\documentclass[preprint,12pt,authornum]{elsarticle}

\usepackage{multirow}
\usepackage{amsmath}
\usepackage{amssymb}
\usepackage{graphicx}
\usepackage{subcaption}
\usepackage{caption}
\usepackage[utf8]{inputenc}
\usepackage{geometry}
\usepackage{xcolor}
\usepackage{enumerate}
\usepackage{enumitem}
\usepackage[ruled, vlined]{algorithm2e}
\usepackage{algpseudocode}
\usepackage{setspace}
 \usepackage{url}
 \usepackage{chngcntr} 
\usepackage[colorlinks = true,
            linkcolor = blue,
            urlcolor  = blue,
            citecolor = blue,
            anchorcolor = blue]{hyperref}
\usepackage{natbib}

\setcitestyle{square, comma, numbers,sort&compress}
\usepackage{tabularray}
\usepackage{soul}

\journal{arXiV}

\begin{document}

\begin{frontmatter}



\title{Review and Validation of Stochastic Ground Motion Models: which one does it better?}




\author[inst1,inst3]{Maijia Su\corref{cor3}}
\author[inst2]{Mayssa Dabaghi}
\author[inst1]{Marco Broccardo\corref{cor1}}

\affiliation[inst1]{organization={Department of Civil, Environmental and Mechanical Engineering, University of Trento, Trento},
            country={Italy}}            
\affiliation[inst2]{organization={Department of Civil and Environmental Engineering,  American University of Beirut, Beirut}, 
            country={Lebanon}}
\affiliation[inst3]{organization={Department of Civil and Environmental Engineering, University of California, Berkeley},
            country={United States}}

\cortext[cor1]{marco.broccardo@unitn.it}
\cortext[cor3]{maijia.su@unitn.it (PhD Candidate), maijiasu@berkeley.edu (Visiting PhD Student)}

\begin{abstract}
Stochastic ground motion models (GMMs) are gaining popularity and momentum among engineers to perform time-history analysis of structures and infrastructures. This paper aims to review and validate \textit{hierarchical} stochastic GMMs , with a focus on identifying their “optimal” configuration. We introduce the word ``\textit{hierarchical}'' as its formulation contains two steps:  (1) selecting a modulated filtered white noise model (MFWNM) to replicate a target record and (2) constructing a joint probability
density function (PDF) for the parameters of the selected MFWNM, accounting for the record-to-record variability.
In the first step, we review the development of MFWNMs and explore “optimal” modeling of time-varying spectral content.  Specifically, we investigate different frequency filters (single- and multi-mode) and various trends (constant, linear, and non-parametric) to describe the filters’ time-varying properties.
In the second step, the joint PDF is decomposed into a product of marginal distributions
and a correlation structure, represented by copula models. We explore two copula models: the
Gaussian copula and the R-vine copula. The hierarchical GMMs are evaluated by comparing
specific statistical metrics, calculated from 1,001 real strong motions, with those derived from their corresponding synthetic dataset.
Based on the selected validation metrics, we conclude that (1) Hierarchical stochastic GMMs can generate ground motions with high statistical compatibility to the real datasets, in terms of four key intensity measures and linear- and nonlinear-response spectra;  (2) A parsimonious 11-parameter MFWNM, incorporating either the Gaussian copula or the R-vine copula, offers sufficient and similar accuracy.
\end{abstract}

\begin{keyword}
Synthetic ground motion  \sep Copula modeling \sep Model validation
\end{keyword}



\end{frontmatter}


\section{Introduction}
\noindent
\label{sec1}
In earthquake engineering,  a proper definition of seismic input is crucial for structural seismic design and seismic risk assessment. The seismic input is typically represented by a suite of  acceleration time series. Engineers often select a suite of original or modified ground motions (GM) records to match a target response spectrum \cite{naeim_selection_2004,wang_ground_2011,iervolino_review_2008}. However, this GM selection method has several well-known limitations. First, the selected GMs will not recur in the future. Second, modified GMs may lack physical significance, potentially containing unnatural frequency content or temporal amplitude variations \cite{grigoriu_scale_2011, luco_does_2007}. Third, the response spectrum is an incomplete measure of GM time series, as it fails to capture spectral nonstationarity and duration \cite{jeong_effect_1988,beck_moving_1993,bradley_correlation_2011}. These shortcomings can lead to an ambiguous and improper representation of GM characteristics and uncertainty, which directly impacts the reliability of building design and seismic risk assessments. 

In alternative to the GM selection methods, stochastic ground motion models (GMMs) are commonly used to simulate synthetic GM time series.  It is critical that stochastically generated GMs capture meaningful engineering characteristics and their (large) variability. As documented in \cite{conte_fully_1997}, stochastic GMMs have evolved from stationary processes  to fully temporal- and spectral-nonstationary processes, originally designed to replicate single historical GMs.  More recently, site-based stochastic GMMs have been developed to account for both the intrinsic uncertainty of individual GMs and the record-to-record variability  \cite{alamilla_evolutionary_2001,pousse_nonstationary_2006,rezaeian_simulation_2012,yamamoto_stochastic_2013,medel-vera_stochastic_2016,vlachos_predictive_2018,dabaghi_simulation_2018,radu_site-specific_2018,sabetta_simulation_2021,wang_stochastic_2024}. This record-to-record variability is related to the uncertainty of the GMM parameters within a specific earthquake scenario or between different scenarios.

The formulation of site-based GMMs contains two critical steps: (\textit{i}) choose a stochastic GMM  to simulate artificial GMs similar to a target record, and (\textit{ii}) construct a joint probability density function (PDF) for the GMM parameters to account for the record-to-record variability.  
In this first step, various types of stochastic GMMs are available, including the modulated filtered white noise model \cite{kubo_simulation_1979,rezaeian_stochastic_2008},  autoregressive moving average model \cite{chang_arma_1982,gersch_time_1985,conte_nonstationary_1992,mobarakeh_simulation_2002}, filtered modulated Poisson pulses model \cite{lin_nonstationary_1965,lin_random_1986,sgobba_evolutionary_2011}, wavelet-based model \cite{lilly_generalized_2012, yamamoto_stochastic_2013,wang_simulation_2022}, and spectral representation of stochastic processes \cite{shinozuka_simulation_1991,liang_simulation_2007,vlachos_multi-modal_2016}, etc.  These stochastic processes include built-in randomness, accounting for the inherent variability in recorded GMs. Essentially, a recorded GM can be viewed as a sample from a stochastic GMM with specific parameters, such as the motion duration, predominant frequency, and expected total energy.  
The second step addresses the limitations of using a single stochastic GMM, which cannot fully capture the natural variability of GMM parameters across different records in certain earthquake scenarios. To model this variability, the current approach contains two steps \cite{alamilla_evolutionary_2001, rezaeian_simulation_2012,yamamoto_stochastic_2013,vlachos_predictive_2018,dabaghi_simulation_2018}: (1) fitting marginal distributions to the parameters of a selected stochastic GMM, and (2) using the Nataf transform to construct the joint PDF by modeling the correlation structure of the GMM parameters in Gaussian space. Next, a site-based GMM require a third step—developing predictive equations that relate the GMM parameters to seismological variables, such as the earthquake magnitude $M$, source-to-site distance $R$, and shear-wave velocity of the superficial soil layers of the site (e.g., $V_{S30}$). This additional step is beyond the scope of this paper and warrants further investigation. 

For convenience, the integration of the two-step formulation is termed \textit{hierarchical stochastic GMMs}. 
These hierarchical models can be fitted to a set of selected GMs believed to reflect the properties of future earthquake scenarios.  Therefore, hierarchical GMMs are data-driven and enable  the identification of models that best match observed records. Once constructed, hierarchical GMMs provide a well-defined probabilistic description of the simulated acceleration time series for the design scenarios. Naturally, they avoid the concern of   GM modification.  As a result, hierarchical GMMs can address the above-mentioned limitations of  the GM selection method. Moreover, they can augment real GM datasets, providing an ``unlimited'' number of compatible records for the selected dataset. Therefore, hierarchical stochastic GMMs enable the effective application of  Uncertainty Quantification in earthquake engineering.

This paper aims to review and validate hierarchical stochastic GMMs, with a focus on identifying their ``optimal" configuration within the two-step formulation. Specifically,  the first step investigates a popular family of stochastic GMMs: the modulated filtered white noise models (MFWNMs). Despite the extensive development of MFWNMs since 1947 \cite{housner_characteristics_1947}, there is a lack of comprehensive comparisons, leaving open questions about which configuration performs best. To address this task, we review the development of MFWNMs, with a comprehensive and concise summary presented in Section \ref{sec_review}.  Based on this, we select representative configurations of MFWNMs, and perform a comprehensive validation and comparison on their performance.

The second step explores the dependence structure among the random GMM parameters. Most GMMs assume Gaussian dependence, typically implemented using the Nataf transform. However, this paper adopts a more general approach by using the R-vine copula model \cite{bedford_vinesnew_2002}. The R-vine copula can capture complex dependencies, such as asymmetric and tail dependence\footnote{Asymmetric dependence refers to unequal correlation strength in one extreme direction (e.g., extremely large values) compared to the other direction (e.g., extremely small values). Neglecting tail dependence implies assuming that extreme events in both upper and lower tails are asymptotically uncorrelated.}, which may affect the prediction of extreme events. 
A recent application of R-vine copulas to stochastic GMMs is found in \cite{peng_stochastic_2023}. 
This study investigates whether such dependencies are critical for simulating GMs.

This paper validates and compares various model configurations using a large dataset comprising 1,001 strong motion records. The model performance is evaluated by comparing specific statistical metrics calculated from the real dataset with those derived from the corresponding synthetic dataset. These statistical metrics correspond to key seismic performance indicators, including four intensity measures: peak ground acceleration (PGA), peak ground velocity (PGV), Arias intensity ($I_a$), and significant duration ($D_{5-95}$)\footnote{The time interval during which the cumulative Arias intensity increases from 5\% to 95\%.}, as well as the response of both linear and nonlinear idealized single-degree-of-freedom (SDOF) structures (i.e., elastic and inelastic response spectra). Various statistical quantities are used for evaluation, including cumulative distribution functions (CDFs), standard deviation, quantiles, and Pearson correlation coefficients.

This paper is structured into two parts. The first part, covered in Sections 2 to 6, focuses on selecting the best configuration of MFWNMs. In Section \ref{sec_review}, we review the MFWNMs and select a set of representative configurations that are used for comparison. In Section \ref{sec2}, we formulate a general framework for MFWNMs using spectral representation, following the study  \cite{broccardo_spectral-based_2017}. In Section \ref{sec3}, we outline how to fit  MFWNMs matched to a historical record. In Section \ref{sec4}, we design and validate the performance of a baseline MFWNM with 11 parameters. In Section \ref{sec4_II}, we present the comparison and validation of 8 different models (including the baseline model) to identify the optimal configuration.  

The second part, presented in Sections \ref{sec5} and Section \ref{sec6}, focuses on selecting the  ``correct" dependence structure. In Section \ref{sec5}, we detail the procedure for constructing two joint PDFs of the GMM parameters, using the same marginals but two different dependence structures: Gaussian dependence and Vine-copula-based dependence. In Section \ref{sec6}, we validate and compare two hierarchical GMMs, which integrate the optimal MFWNM identified in Section \ref{sec4_II} with the two joint PDFs formulated in Section \ref{sec5}.

\section{Review of Modulated Filtered White Noise Models }
\label{sec_review}
This section presents a comprehensive and concise review of MFWNMs. For further details, readers can refer to review papers such as \cite{liu_synthesis_1970, shinozuka_stochastic_1988, kozin_autoregressive_1988, conte_fully_1997, rezaeian_stochastic_2008}. The evolution of MFWNMs began with simple white noise processes (Housner, 1947 \cite{housner_characteristics_1947}; Bycroft, 1960 \cite{ bycroft_white_1960}), progressed to stationary filtered white noise models like the Kanai-Tajimi filter (Kanai, 1957 \cite{k_semi-empirical_1957} and Tajimi, 1960 \cite{tajimi_statistical_1960}, hereafter referred to as KT-filter), and eventually evolved into fully temporal- and spectral-nonstationary random processes (e.g., \cite{kubo_simulation_1979, conte_fully_1997}). There are two approaches to simulate non-stationary stochastic processes: (1) in the time domain by solving a linear filter excited by Gaussian white noise \cite{rezaeian_stochastic_2008}, or (2) in the frequency domain using spectral representations based on Priestley’s evolutionary theory \cite{shinozuka_simulation_1991,liang_simulation_2007,broccardo_spectral-based_2017}. These two simulation approaches are mathematically equivalent and have been numerically validated in  \cite{su_importance_2024}. This review summarizes how this family of stochastic models characterizes key aspects of natural GMs, including time-varying amplitude, spectral nonstationarities, and multi-mode spectral characteristics. Finally, we introduce the fitting procedure to recorded time series.

The time-varying amplitudes of  GM acceleration are typically represented by a deterministic function, known as the time-modulating function. This function usually comprises three phases: an initial build-up with low intensity, a strong phase with maximum amplitude in the middle, and a final decay phase as the shaking subsides. In the literature, various functional shapes have been proposed for the time-modulating function. Here is a list of examples: 
exponential function (Bolotin, 1960 \cite{bolotin_statistical_1960}),
time-modulated harmonics (Bogdanoff et al., 1961  \cite{bogdanoff_response_1961}), 
piece-wise modulating functions (Housner and Jennings, 1964 \cite{housner_generation_1964}; Amin and Ang, 1968 \cite{amin_nonstationary_1968}), 
double-exponential function (Shinozuka and Sato, 1967 \cite{shinozuka_simulation_1967}), 
Gamma function (Saragoni and Hart,  1973 \cite{rodolfo_saragoni_simulation_1973}),
Beta function (Arias et al., 1976 \cite{arias_approximate_1976}), 
piecewise linear function (Der Kiureghian and Crempien, 1989 \cite{der_kiureghian_evolutionary_1989}), 
log-normal distribution (Stafford et al., 2009 \cite{stafford_energy-based_2009}), 
a multi-peak envelope function (Ghasemi and Ashtari, 2014 \cite{ghasemi_combinatorial_2014}),
spline interpolation of GM cumulative energy  (Broccardo and Dabaghi, 2018 \cite{broccardo_spectral-based_2017}), 
and equivalent average-intensity envelope (Wang et al., 2021, \cite{wang_new_2021}). 
Note that the spline-modulating function \cite{broccardo_spectral-based_2017} is non-parametric and flexible to capture multi-peak amplitudes, as such, in this study, we adopt this formulation.  

The spectral nonstationarities arise from the arrival times of different seismic wave types (i.e., P-waves, S-waves, and surface waves) and complex effects of wave propagation. In GM simulations, there are several strategies for modeling these time-varying spectral contents. One approach involves building a\textit{ frequency-nonstationary} process by combining a series of time-modulated \textit{frequency-stationary} processes. This combination can be applied in either the time domain (i.e., combining multiple adjacent locally stationary time segments, e.g., \cite{rodolfo_saragoni_simulation_1973, muscolino_generation_2021}) or the frequency domain (i.e., combining multiple time-modulated stationary processes defined on non-overlapping frequency sub-bands, e.g., \cite{der_kiureghian_evolutionary_1989, conte_fully_1997, wang_evolutionary_2018}).
Another popular technique uses stochastically-excited filters with parameters that slowly change over time. These models rely on characterizing the time-varying trend of the filter parameters. For instance, an early effort by Kubo and Penzien (1979) \cite{kubo_simulation_1979} modified the frequency-invariant KT-filter by introducing a time-dependent dominant frequency. Subsequent studies have proposed various trend functions for the two key KT-filter parameters—the damping ratio $\zeta_g$ and the filter frequency $\omega_g$. An incomplete list includes:
\begin{itemize}
    \item third-order polynomials for $\zeta_g$ and $\omega_g$ (Deodatis and Shinozuka, 1988 \cite{deodatis_autoregressive_1988});
    \item a constant $\zeta_g$ and an bi-exponentially decaying $\omega_g$ (Fan and Ahmadi, 1990 \cite{fan_nonstationary_1990});
    \item a linearly varying $\zeta_g$ and exponentially decaying $\omega_g$ (Beck and Papadimitriou, 1988 \cite{beck_moving_1993});
    \item a constant $\zeta_g$ and a linear function for $\omega_g$ (Rezaeian and Der Kiureghian, 2008 \cite{rezaeian_stochastic_2008});
    \item both linear functions for $\zeta_g$ and $\omega_g$ (Sgobba et al, 2011 \cite{sgobba_evolutionary_2011}).
\end{itemize}
Additionally, instead of being defined in the time domain, the trend function can be formulated in the energy domain \cite{vlachos_multi-modal_2016}. Beyond these two primary groups, alternative methods for accounting for frequency nonstationarity can be found in \cite{lin_evolutionary_1987, yeh_modeling_1990, grigoriu_mexico_1988}.

Multi-mode spectral models, which account for the presence of multiple dominant frequencies in recorded GMs, have also gained attention. The early attempt can be traced to the work (Liu and Jhaveri, 1969 \cite{liu_spectral_1969}), which used a sum combination of two single-mode frequency transfer functions. Subsequent research \cite{faravelli_stochastic_1988,deodatis_autoregressive_1988, vlachos_multi-modal_2016} formulated the multi-mode shapes using a weighted sum of multiple single-mode KT filters. A recent study  \cite{waezi_simulation_2022} introduced a broad frequency spectrum model by combining high- and low-frequency filters. Provided with this large literature, this study explores the ``optimal'' models of the time-varying frequency content, specifically using and testing various frequency filters (single- and multi-mode) and trend functions (constant, linear, and non-parametric), as introduced in Section \ref{sec2}.

The limitation of the KT filter to realistically capture low-frequency energy of recorded GMs is well recognized. First of all, a finite value of the Power Spectral Density (i.e., the KT filter) at zero frequency makes the simulated acceleration process non-integrable.  To address this issue, Clough and Penzien \cite{clough_dynamics_1993} modified the KT filter by introducing a damped second-order filter, which \textit{acts like} a high-pass filter to ensure zero energy at zero frequency.  The need for high-pass filtering in the KT model was also recognized by Beck and Papadimitriou \cite{beck_moving_1993}, aiming to mimic Brune’s source model \cite{brune_tectonic_1970}.  Brune’s model is a simplified source model that describes the frequency distribution of energy released from earthquake ruptures. The high-pass filtering can be realized using critical-damped linear oscillators \cite{beck_moving_1993} or Butterworth filters \cite{vlachos_multi-modal_2016}. These filters have one common parameter, the corner frequency $f_c$, which was commonly regarded as a fixed parameter \cite{conte_fully_1997,rezaeian_simulation_2012} or correlated to semiological variables, such as earthquake moment \cite{dabaghi_simulation_2018} and earthquake duration \cite{vlachos_multi-modal_2016}. 
However, a recent paper \cite{su_importance_2024} pointed out that the high-pass filtering in stochastic GMMs represents a more complex process, accounting for both physical effects (i.e., earthquake source properties, wave attenuation through the earth's crust, and local soil amplification) and artificial effects (i.e., noise processing techniques applied to raw recorded GMs). The study \cite{su_importance_2024} emphasized that $f_c$ should be a free parameter and proposed optimizing $f_c$ such that the low-frequency content of MFWNMs accurately matches that of recorded GMs.

In the fitting procedure, most of the proposed stochastic GMMs separately fit the time-modulating functions and frequency-modulating functions (i.e., time-varying filters). 
For time-modulating functions, a common approach is to match the cumulative energy of the model to that of a target accelerogram. The mismatch is typically quantified using mean square errors \cite{yeh_modeling_1990, vlachos_multi-modal_2016} or by comparing specific features (e.g., $I_a$, $D_{5-95}$) of the cumulative energy curves \cite{alamilla_evolutionary_2001, rezaeian_stochastic_2008}.
For frequency-modulating functions, the fitting procedures vary depending on how spectral nonstationarity is measured. The relevant GMM parameters are then estimated to match these computed nonstationary metrics. Spectral nonstationarity can be quantified using the coefficients of time-varying autoregressive and moving averaging models (ARMA) \cite{beck_moving_1993, conte_nonstationary_1992, mobarakeh_simulation_2002, sgobba_evolutionary_2011}. Analytical formulations exist to relate ARMA coefficients to the parameters of the KT filter \cite{chang_arma_1982, conte_nonstationary_1992}.
Additionally, nonstationary metrics can be quantified using specific characteristics of GM time series. For example, the mean zero-level up-crossing rate can capture the predominant frequency, while the rate of negative maxima or positive minima characterizes the bandwidth \cite{rodolfo_saragoni_simulation_1973, yeh_modeling_1990, rezaeian_stochastic_2008, medel-vera_stochastic_2016, waezi_evolutionary_2017}.
Finally, frequency nonstationarity can be directly computed using an empirical or pseudo evolutionary power density spectrum (EPSD). Various techniques are employed to compute the EPSD of recorded GMs, including short-time Thomson’s multiple-window spectrum \cite{conte_fully_1997, vlachos_multi-modal_2016, broccardo_spectral-based_2017}, wavelet transform \cite{spanos_evolutionary_2004}, Gabor transform \cite{kumar_timefrequency_2016}, Hilbert spectral analysis \cite{zhang_generation_2012, bagheri_simulation_2014}, and S-transform \cite{cui_timefrequency_2020}. These methods measure how frequency content evolves over time in GMs.

\section{Parameterization of Stochastic Ground Motion Models}

\noindent
\label{sec2}
This section introduces the stochastic GMM used for simulating far-field non-pulse-like  GMs. The model is composed of four parametric modules, each representing key physical characteristics of natural GMs. Specifically, we adopt the stochastic GMM \cite{broccardo_spectral-based_2017} and extend it by enriching the parametric options within each module. First, we present the general framework of this GMM, followed by a detailed description of each module.

\subsection{General framework}
\label{sec2_1}
This subsection presents a general framework for parametric stochastic GMMs based on spectral representation. These GMMs build on a modulated filtered white noise process in the frequency domain, serving as the frequency-domain counterpart to the time-domain formulation proposed by Rezaeian and Der Kiureghian \cite{rezaeian_stochastic_2008}. The synthesis of GM accelerations through a spectral-based procedure involves four steps: 
  \begin{itemize}
      \item \textbf{\textit{Step 1:} Definition of Evolutionary Power Spectral Density (EPSD).} \\
          Consider a stochastic process $A(t)$ with one-sided EPSD $S_{AA}(t,\omega)$ defined by:
      \begin{equation}
          \label{sec2_eq1}
          S_{AA}(t,\omega) = q^2(t|\boldsymbol{\theta}_{T}) \phi(\omega|\boldsymbol{\theta}_{\phi}(t)),
      \end{equation}
          where $q(t|\boldsymbol{\theta}_{T})$ represents a parametrized time-modulating function with parameters  $\boldsymbol{\theta}_{T}$, and $\phi(\omega|\boldsymbol{\theta}_{\phi}(t))$ is a frequency-modulating function with time varying parameters $\boldsymbol{\theta}_{\phi}(t)$.
             
          The time-modulating function describes the temporal energy evolution of simulated GMs, i.e., $q^2(t|\boldsymbol{\theta}_{T})=\int_{0}^{+\infty}S_{AA}(t,\omega)d\omega$. The frequency-modulating function defines the instantaneous spectral energy distribution; moreover, at each instant of time $\int_0^{+\infty} \phi(\omega|\boldsymbol{\theta}_{\phi}(t)) d\omega = 1$. This allows to separate the spectral nonstationarity,  controlled by $\phi(\omega|\boldsymbol{\theta}_{\phi}(t))$, from the temporal nonstationarity controlled by $q(t|\boldsymbol{\theta}_{T})$.
  The definition of  $q(t|\boldsymbol{\theta}_{T})$ is detailed in  Section \ref{sec2_2} and the definition of $\phi(\omega|\boldsymbol{\theta}_{\phi}(t))$ is detailed in Section \ref{sec2_3} and  Section \ref{sec2_4}.
  
      \item \textbf{\textit{Step 2:}  Simulation of Temporal- and Spectral Non-Stationary Random Process.}\\
       Simulate a fully temporal- and spectral-non-stationary random process using the discrete version of spectral representation \cite{shinozuka_stochastic_1988}:
            \begin{equation}
               \label{sec2_eq2}
               A(t) = \sum_{k=1}^{K}\sigma_{A}(t,\omega_k) [ Z_{k}\sin(\omega_k t)+Z_{K+k}\cos(\omega_k t)], \  t\in[0,t_f].
            \end{equation}           
Here,  $t_f$ represents the time length of the simulated process. The EPSD is discretized at equally spaced frequencies, i.e.,  $\omega_k = (k-1) \Delta \omega$, $\Delta\omega$ is the discretization step, and $\Omega_{u} = (K-1)\Delta\omega$ is the upper cut-off frequency.\footnote{This study sets $K=\lceil t_f/dt \rceil \in \mathbb{Z^+}$ and fixes $\Omega_{u}$ as 25Hz, where the simulation time step is chosen as $dt=0.02$s. This setting indicates that the
 simulated process is non-periodic, since it statifies $2\pi/\Delta \omega \geq t_f$ \cite{shinozuka_stochastic_1988}.
} 
The vector $\boldsymbol{Z} = \{Z_k|k=1,2,...,2K\}$ is composed by i.i.d. standard normal random variables. The standard deviation of the Fourier series amplitude in Eq.\eqref{sec2_eq2} is defined as $\sigma_{A}(t,\omega_k) = \sqrt{\bar{S}_{AA}(t,\omega_{k})\Delta\omega}$, where $\bar{S}_{AA}(t,\omega_k) = {S}_{AA}(t,\omega_k)/\sum_{k'=1}^K (\phi(\omega_{k'}|\boldsymbol{\theta}_{\phi}(t))\Delta\omega )$ is adjusted to ensure energy consistency in the discrete representation.  
      \item \textbf{\textit{Step 3:}  High-Pass Filtering.} \\
         Convolve ${A}(t)$ with a high-pass filter $h_f(t|f_c)$ to obtain 
         \begin{equation}
              \label{sec2_eq2+}
              D(t) =  A(t) * h_f(t|f_c), 
         \end{equation}
         where the symbol * denotes a convolution operator. The parameter, corner frequency $f_c$, determines the point where the low-frequency content of simulated GMs starts to decrease sharply. In this study, the output process $D(t)$ is a displacement process \cite{beck_moving_1993,rezaeian_stochastic_2008}; then, the filtered acceleration process is derived by differentiating each realization of $D(t)$. We name this high pass filtered acceleration process as $A'_g(t)$.


The high-pass filtering  (detailed in Section \ref{sec2_5}) ensures zero residual velocity and displacement at the end of the simulated GMs. Notice that since $f_c$ controls the low-frequency content of simulated GMs, it can be fitted to match this region of interest. 
      \item \textbf{\textit{Step 4:} Energy Correction.}\\
          The high pass filter in Step 3 alters the energy content of the process. Therefore, to ensure consistent Arias intensity \cite{dabaghi_simulation_2018}, we multiply each $a'_g(t)$ (i.e., realization of $A'_g(t)$) with an energy-correction factor $\kappa_E\geq 1$ and obtain the simulated GM:
           \begin{equation}
               \label{sec2_eq2++}
               a_g(t) = \kappa_{E}a'_g(t). 
           \end{equation}   
Since $\kappa_{E}$ is a fixed value for each realization $a'_g(t)$,  then $a_g(t)$  is a sample of the process $A_g(t)=\kappa_{E}A'_g(t)$.
           The value of $\kappa_E$ is determined by the specific choices of the parametric models (see Section \ref{sec2_2} to \ref{sec2_5}) and their estimated model parameters. The detailed formula for computing $\kappa_E$ is introduced in Section \ref{sec2_5}. 
  \end{itemize}

%
%
%

In summary, the spectral-based stochastic GMM integrates four parametric sub-models: $q(t|\boldsymbol{\theta}_T)$, $\phi(\omega|\boldsymbol{\theta}_{\phi}(t))$, $\boldsymbol{\theta}_{\phi}(t|\boldsymbol{\theta}_F)$  and $h_f(t|f_c)$. By assembling these sub-models within the 4-step procedure, we treat the GM time series generation model as a stochastic simulator:
\begin{equation}
\label{sec2_eq2+++}
  A_g(t) = \mathcal{M}(t|\boldsymbol{\theta},\boldsymbol{Z}), t \in [0,t_f]
\end{equation}
Here, $\boldsymbol{\theta} = [\boldsymbol{\theta}_T, \boldsymbol{\theta}_F,f_c] \in \mathcal{D}_{\boldsymbol{\theta}}$ denotes a finite number of engineering meaningful GM parameters associated with duration, frequency content characteristics, and cumulative energy of the acceleration process. While $\boldsymbol{\theta}$ is a deterministic input vector, the process $A_g(t)$ is stochastic due to the aleatory uncertainty of the Gaussian white noise $\boldsymbol{Z}$. In this study, each recorded GM time series is regarded as one sample drawn from $A_g(t)$, with corresponding fitted parameters $\boldsymbol{\hat\theta}$. To account for epistemic variability arising from different records (as done in \cite{rezaeian_simulation_2012,vlachos_predictive_2018}), Section \ref{sec4_II} introduces a second layer of uncertainty to ${\boldsymbol{\theta}}$.

\subsection{Time-modulating function  $q(t|\boldsymbol{\theta}_T)$}
\label{sec2_2}
The time-modulating function describes the variation in amplitude of the ground motion (GM) acceleration process over time. It can be shown that this function directly controls the buildup of the expected Arias intensity of the process, i.e.
\begin{equation}
\label{sec2_2_eq2}
 \hat{I}_{\mathrm{a}}(t|\boldsymbol{\theta}_T)  = \mathbb{E} \left[ \frac{\pi}{2g}\int_0^t A^2(\tau)d\tau   \right] = \frac{\pi}{2g} \int_0^t q^2(\tau|\boldsymbol{\theta}_T)d\tau,
\end{equation}
where $g$ is the gravitational acceleration. The curve $\hat{I}_{\mathrm{a}}(t|\boldsymbol{\theta}_T)$ is used to match to the cumulative Arias Intensity, ${I}_{\mathrm{a}}(t)$, of a recorded GM (also known as the Husid plot). 
%

Previous GMMs (e.g., \cite{housner_generation_1964,shinozuka_simulation_1967,amin_nonstationary_1968,rodolfo_saragoni_simulation_1973,stafford_energy-based_2009,rezaeian_simulation_2012}) have used various functional forms to define $q(t|\boldsymbol{\theta}_T): [0,t_f] \rightarrow \mathbb{R}_0^+$.  Most time-modulating functions are parametric, meaning the shape of the evolving  GM amplitude is constrained by the selected function form and the parameters are found by minimizing the mean square error between  $\hat{I}_{\mathrm{a}}(t|\boldsymbol{\theta}_T)$ and $I_{\mathrm{a}}(t)$. This study uses a simple non-parametric cubic Hermite spline interpolator (which is a non-decreasing interpolator), as introduced  in \cite{broccardo_spectral-based_2017}. 
%
%
%
In this setting, the GM duration $[0,t_f]$ is partitioned to discrete intervals at selected time points   $t_{p_i},i=1,2,...,n_p$, where $t_{p_i}$ denotes the time when the process reaches $p_i\%$ of the total cumulative energy $I_a=I_a(t_{100})$ (also termed as Arias intensity in the literature \cite{arias1970measure}). The Hermite spline is constrained to points $\{(t_{p_i},I_a(t_{p_i}))|i=1,2,...,n_p\}$ and a cubic polynomial is built in each sub-interval $[t_{p_i},t_{p_{i+1}}]$. This study chooses the same interpolation points as in \citep{broccardo_preliminary_2019}, which sets  $n_p=7$ and  $p_i = [0,5,30,45,75,95,100]$.  In this case, with fixing $t_{0} = 0$ and defining sub-interval time length $D_{p_i-p_{i+1}} = t_{p_{i+1}} - t_{p_i}$, the spline modulating function comprises of 7 parameters, denoted as $ \boldsymbol{\theta}_{T} = [D_{0-5},D_{5-30},D_{30-45},D_{45-75}, D_{75-95},D_{95-100},I_a]$. When these parameters are computed, the first derivative of  $\hat{I}_{\mathrm{a}}(t|\boldsymbol{\theta}_T)$ is continuous and corresponds to $\frac{\pi}{2g} q^2(t|\boldsymbol{\theta}_T)$. Note that the defined intervals (including combinations of them) are engineering-meaningful quantities that describe the energy evolution of the GM.
Therefore, the ``non-parametric'' spline modulating function can be viewed as a parametric model (with parameters $\boldsymbol{\theta}_{T}$). However, these parameters are determined without using optimization process. Moreover, this approach offers flexibility in shaping the GM amplitude evolution without being constrained by a specific functional form. For instance, it can accommodate recorded GMs with multiple strong-motion phases.

\subsection{Frequency-filter function $\phi(\omega|\boldsymbol{\theta}_{\phi}(t))$}
\label{sec2_3}

A time-varying frequency modulating function is defined as $\boldsymbol{\theta}_{\phi}(t): [0,t_f] \times \mathbb{R}^+_0 \rightarrow \mathbb{R}_0^+$. Typically, this function is chosen to be parametric, $\phi(t,\omega) = \phi(\omega|\boldsymbol{\theta}_\phi(t))$, with time-varying filter parameters $\boldsymbol{\theta}_\phi(t)$.
%
%
%
%
This subsection introduces different functional forms for the filters $\phi(\omega|\boldsymbol{\theta}_{\phi}(t))$. The next subsection introduces the modeling of the time-varying trend for parameters  $\boldsymbol{\theta}_{\phi}(t)$.

\subsubsection{Second-order filter of linearly under-damped oscillator}
\label{sec2_3_1}
This filter is defined as the pseudo-acceleration frequency response function of an under-damped linear-elastic SDOF oscillator. This frequency filter is essentially the frequency-domain equivalent of the time-domain filter used in \cite{rezaeian_stochastic_2008}.  It is formulated as follows:  
\begin{equation}
\label{filter_II}
    \phi(\omega|\boldsymbol{\theta}_\phi(t)) = \phi_0(t) \frac{\omega_g^4(t)}{(\omega_g^2(t)-\omega^2)^2+4\zeta_g^2(t)\omega^2_g(t)\omega^2},
\end{equation} 
where $\boldsymbol{\theta}_{\phi}(t) = [\omega_g(t),\zeta_g(t)]$ with $\omega_g(t)$ as the predominant frequency of the filter and  $\zeta_g(t)$ as the filter bandwidth. $\phi_0(t)$ is a normalizing constant introduced to guarantee a unit area under the filter at any time, i.e., $\int_0^{+\infty} \phi(\omega|\boldsymbol{\theta}_{\phi}(t)) d\omega = 1$. The definition of $\phi_0(t)$ is also the same for the following frequency filters.  This filter is one-time-differentiable, which is necessary for a time-domain formulation of GMs. Moreover, it is not integrable, due to non-zero values at zero frequency. This issue can be addressed either using the Clough-Penzien filter  \cite{clough_dynamics_1993} (which is not necessary for this study) or the high-pass filter (as introduced in Eq. \eqref{sec2_eq2+}).

\subsubsection{Kanai-Tajimi filter}
\label{sec2_3_2}
The Kanai-Tajimi filter \cite{k_semi-empirical_1957,tajimi_statistical_1960} is widely used in earthquake engineering. Its formulation with time-varying parameters is given by:
\begin{equation}
\label{filter_KT}
    \phi(\omega|\boldsymbol{\theta}_\phi(t)) = \phi_0(t) \frac{\omega_g^4(t) + 4\zeta_g^2(t)\omega_g^2(t)\omega^2}{(\omega_g^2(t)-\omega^2)^2+4\zeta_g^2(t)\omega^2_g(t)\omega^2},
\end{equation}
where it shares the same filter parameters as the second-order filter. This filter is both not integrable and differentiable; therefore, in the continuous version, it cannot be used in the time-domain formulation proposed by Rezaeian and Der Kiureghian \cite{rezaeian_stochastic_2008}.

\subsubsection{Convex combination of single-mode filters}
\label{sec2_3_3}
We investigate a mixed filter defined as a linear combination of uni-modal filters. This mixed filter is designed to capture a multi-mode spectral shape, defined as a convex combination of $J$-individual single-mode filters:
\begin{equation}
\label{filter_Convex_II}
    \phi(\omega|\boldsymbol{\theta}_\phi(t)) = \phi_0(t) \sum\limits_1^J \pi_j(t)\phi(\omega|\boldsymbol{\theta}_{\phi,j}(t)),
\end{equation}
where, in this study,  $\phi(\omega|\boldsymbol{\theta}_{\phi,j}(t))$ is chosen as either the  Kanai-Tajimi filter (Eq. \eqref{filter_KT}) or the second-order filter (Eq. \eqref{filter_II}). Hence, all  the parameters of the mixed filter are grouped as $\boldsymbol{\theta}_{\phi}(t) = [\omega_{g,j}(t),\zeta_{g,j}(t),\pi_j(t)|j=1,2,...,J]$, with $[\omega_{g,j}(t),\zeta_{g,j}(t),\pi_j(t)]$ as the parameters of the $j$-th  single-mode filter. The weight coefficients are positive and constrained by $\sum\limits_1^J \pi_j(t) = 1$, thus this mixture filter has $3J-1 $ free parameters.

\subsubsection{Cascade combination of single-mode filters}
\label{sec2_3_4}
Similar to the mixed filter in the last subsection, we explore a new filter, defined as the multiplication of  $J$ single-mode filters:
\begin{equation}
\label{filter_cascade_II}
    \phi(\omega|\boldsymbol{\theta}_\phi(t)) = \phi_0(t) \prod\limits_1^J  \phi(\omega|\boldsymbol{\theta}_{\phi,j}(t)),
\end{equation}
where the filter contains $2J$ free parameters, i.e., $\boldsymbol{\theta}_{\phi}(t) = [\omega_{g,j}(t),\zeta_{g,j}(t)|j=1,2,...,J]$. This filter also can capture the multi-mode spectral shape but with fewer parameters. 

\subsection{Time-varying filter parameters $\boldsymbol{\theta}_\phi(t|\boldsymbol{\theta}_F)$}
\label{sec2_4}

This subsection explores various parametric trends used to characterize the time-varying behavior of the GM parameter vector $\boldsymbol{\theta}_\phi(t)$. The selection of the parametric trend for each scalar element ${\theta}_\phi(t)$ in $\boldsymbol{\theta}_\phi(t)$ is independent of the type of filter function chosen. Generally, the importance of a GM parameter dictates the need for more flexible trends. For instance, seismic responses are more sensitive to the filter frequency $\omega_g(t)$ than to $\zeta_g(t)$, meaning that $\omega_g(t)$ requires greater flexibility. While a more flexible trend can capture complex temporal variations, it also requires a larger number of trend parameters. This study examines three types of parametric trends: constant, linear, and non-parametric. The trend parameters for a chosen filter function $\phi(\omega|\boldsymbol{\theta}_\phi(t))$ fully determine the spectral nonstationarity, and these parameters are included in $\boldsymbol{\theta}_F$.

\subsubsection{Constant trends with time}
\label{sec2_4_1}
The constant trend chooses a typical value at the strong-shaking phase to represent the time-varying filter parameter. That is 
\begin{equation}
\label{sec2_4_eq1}
  \theta_\phi(t) \equiv \theta_{\phi,t_{mid}},
\end{equation}
where $t_{mid}$ is the time of the middle of the strong-shaking phase, taken here as the time of reaching 45\% of the Arias intensity \cite{rezaeian_simulation_2012}. 
In this study, the constant trend is used for the filter bandwidth $\zeta_g(t)$. This choice aligns with previous findings \cite{rezaeian_simulation_2012,broccardo_preliminary_2019}, which concluded that setting the bandwidth as a constant is sufficient.

\subsubsection{Linear trends with time}
\label{sec2_4_2}
Next, a linear function over the interval $[t_5,t_{95}]$ is used to describe the temporal variation:
\begin{equation}
\label{sec2_4_eq2}
    \theta_\phi(t) = \left\{ \begin{matrix}
                   & \theta_\phi(t_{5}) ,&  t_0 \leq t < t_{5}  \\ 
                   & \theta_{\phi,t_{mid}} + \theta'_{\phi,t_{mid}}(t-t_{mid}) ,& t_5 \leq t < t_{95}   \\ 
                   & \theta_\phi(t_{95}) ,&  t_{95} \leq t \leq t_{100}  \\ 
                      \end{matrix} \right..
\end{equation}
Again here, $t_{mid}$ is taken as $t_{45}$, and $\theta_{\phi,t_{mid}}$ has the same physical interpretation as the constant trend model. The slope $\theta'_{\phi,t_{mid}}$ denotes the rate of change of the parameter at $t_{45}$. The prediction outside the $[t_5,t_{95}]$ is assumed constant, avoiding undesired and unphysical estimation of the filter parameters.

\subsubsection{Non-parametric trends with time}
\label{sec2_4_3}
This study also proposes a non-parametric trend based on an interpolation scheme:
\begin{equation}
\label{sec2_4_eq3}
    \theta_\phi(t) = \left\{ \begin{matrix}
                   & \theta_\phi(t_{5}) ,&  t_0 \leq t < t_{5}  \\ 
                   & \sum\limits_{i=1}^{n_I}\varphi_{i}(t)\theta_\phi(t_{p_i}) ,& t_5 \leq t < t_{95}   \\ 
                   & \theta_\phi(t_{95}) ,&  t_{95} \leq t \leq t_{100}  \\ 
                      \end{matrix} \right.
\end{equation}
where $(t_{p_i},\theta_\phi(t_{p_i}))$ is the interpolation point, and $\varphi_i(t)$ is a shape function satisfying conditions: $\varphi_i(t_{p_i})=1$ and $\varphi_i(t_{p_j})=0$ for $j \neq i$. The interpolation occurs within the interval $[t_5,t_{95}]$, ensuring an accurate description of the critical phase of the GM. Extrapolation beyond the defined region is assumed to be constant. We use the same discrete time points $t_{p_i}$ as in the spline modulating function, excluding the first and last point, i.e., $n_I=5$ and  $p_i = [5,30,45,75,95]$. This choice helps to reduce the total number of GM parameters and still capture the transition of the arrival of P-waves, S-waves, and surface waves \cite{broccardo_spectral-based_2017}.  Therefore, when the spline modulating function is used, this trend function includes five parameters, i.e, $[\theta_\phi(t_{5}),\theta_\phi(t_{30}),\theta_\phi(t_{45}),\theta_\phi(t_{75}),\theta_\phi(t_{95})]$ for one scale element of vector $\boldsymbol{\theta}_\phi(t)$. 

This study uses local linear shape functions, leading to a simple polyline interpolation. Other local basis functions can be used without altering the proposed formulation. To avoid abrupt changes at interpolation points, we smooth the polyline by applying a moving average with a 2-second window. 
This non-parametric trend function shares the advantages of the spline modulating functions, including an optimization-free fitting procedure, a flexible ability to capture the temporal variation, and a set of direct physical parameters with engineering significance.

\subsection{High-pass filtering and energy correction}
\label{sec2_5}
As described in Eq.\eqref{sec2_eq2+}, this study uses a critically damped oscillator to filter the process $A(t)$. The filter model  is selected as the impulse displacement response function of this oscillator, denoted as \(h_f(t|f_c) = t\exp(-2\pi f_{c}t)\), where \(f_c\) is the corner frequency. Other, more sophisticated high-pass filters, such as the Butterworth filter, can be used; however, the specific type of high-pass filter is generally not critical. Instead, fitting the corner frequency is considered more important, as similarly shown in the noise processing of raw records \cite{boore_processing_2005}. Therefore, we employ the single-parameter filter introduced earlier. 

The high-pass filtering does not impact the predominant frequency content of the process but partially excludes low-frequency content, leading to a decrease in the total energy. This energy bias is inherently present in the synthetic process $A'_g(t)$. To account for this effect, an energy correction factor $\kappa_E$, introduced in Eq.\eqref{sec2_eq2++}, is used to to linearly scale $A'_g(t)$:  
\begin{equation}
\label{sec2_5_eq1}
\kappa_{E} = \sqrt{\frac{  \mathbb{E} \left[ \frac{\pi}{2g} \int_0^{t_f} A^2(\tau)d\tau \right] }  {\mathbb{E} \left[ \frac{\pi}{2g} \int_0^{t_f} A'^2_g(\tau)d\tau \right]} }.
\end{equation}
Here, the scale factor $\kappa_E$ is defined as the square root of the ratio between the expected Arias intensity of $A(t)$ and $A'_g(t)$.  In Eq.\eqref{sec2_5_eq1}, the term in the numerator is computed using Eq.\eqref{sec2_2_eq2}, and the formula for computing the denominator is derived as follows:
\begin{equation}
\label{eq_sec2_7}
\begin{aligned}
   D(t) & = A(t) * h_f(t|f_c) \\ 
          & = \sum\limits_{k=1}^{K}{{{Z}_{k}}\underbrace{\left( {{\sigma }_{A}}(t,{{\omega }_{k}})\sin ({{\omega }_{k}}t) \right)*{{h}_{f}}(t|{{f}_{c}})}_{{{s}_{k}}(t)}} \\ 
          & +\sum\limits_{k=1}^{K}{{{Z}_{K+k}}\underbrace{\left( {{\sigma }_{A}}(t,{{\omega }_{k}})\cos ({{\omega }_{k}}t) \right)*{{h}_{f}}(t|{{f}_{c}})}_{{{c}_{k}}(t)}}
\end{aligned},
\end{equation}
where $s_k(t)$ and $c_k(t)$ are deterministic functions. Note that $A'_g(t) = \ddot{D}(t)$,  $\mathbb{E}[Z^2_k]=1$ for $k\in[1,2K]$, and $\mathbb{E}[Z_{k}Z_{k'}]=0$ for $k \neq k'$, thus the expected Arias intensity of $A'_g(t)$ is written as:
\begin{equation}
\label{eq_sec2_8}
   \mathbb{E} \left[ \frac{\pi}{2g} \int_0^{t_f}A'^2_g(\tau)d\tau \right] = \frac{\pi}{2g}  \sum_{k=1}^{K} \left( \int_0^{t_f}\ddot{s}_k^2(\tau)d\tau +  \int_0^{t_f}\ddot{c}_k^2(\tau)d\tau  \right),
\end{equation}
where $\ddot{s}_k(t)$ and $\ddot{c}_k(t)$ can be  calculated by their definitions through numerical convolution and differentiation. 


\section{Identification of Stochastic GMM Parameters}
\noindent
\label{sec3}
The spectral-based stochastic GMMs, presented in Section \ref{sec2}, include three sets of parameters: $\boldsymbol{\theta}_T$ of  the time-modulating function $q(t|\boldsymbol{\theta}_T)$, $\boldsymbol{\theta}_F$ of  the frequency modulating function $\phi(t,\omega|\boldsymbol{\theta}_F)$, and  $f_c$ of the high-pass filter $h_f(t|f_c)$. This study uses the same procedures proposed by \cite{rezaeian_simulation_2012,broccardo_spectral-based_2017} to estimate the first two sets of parameters, and the procedure outlined in \cite{su_importance_2024} to estimate $f_c$.

\subsection{Parametric fitting to $q(t|\boldsymbol{\theta}_T)$}
\label{sec3_1}

The parameters $ \boldsymbol{\theta}_{T} = [D_{0-5},D_{5-30},D_{30-45},D_{45-75}, D_{75-95},D_{95-100},I_a]$ of the spline modulating function are determined by their definition. These parameters are physically interpretable and can be  directly computed from the Husid plot $I_a(t)$. As $I_a(t)$ is continuous and non-decreasing, the parameters $t_{p_i}$ are simply the time at the $p_i$ percentiles of this function. 
%
The  time length of each sub-interval is naturally obtained: $D_{p_i-p_{i+1}} = t_{p_{i+1}} - t_{p_i}$.


\subsection{Parametric fitting to $\phi(t,\omega|\boldsymbol{\theta}_F)$}
\label{sec3_2}
The fitting of parameters $\boldsymbol{\theta}_F$ includes two steps: (1) estimate the time-varying filter parameters $\boldsymbol{\hat\theta}_{\phi} (t)$ at each discrete time $t$ by minimizing the discrepancy between the analytical function $\phi(\omega|\boldsymbol\theta_{\phi}(t))$ and a estimated $\hat\phi(t,\omega)$, and (2) estimate parameters $\hat{\boldsymbol\theta}_F$ according to the trend functions $\boldsymbol{\theta}_{\phi} (t|{\boldsymbol\theta}_F)$ selected in Section \ref{sec2_4}.


The first step starts with computing an empirical discrete EPSD  $\hat{S}_{AA}(t,\omega)$ of a target recorded GM. While there are various techniques available to estimate $\hat{S}_{AA}(t,\omega)$, we use the short-time Thomson's multiple-window (STTMW) spectrum because it is not limited by the trade-off between bias and spectral leakage \cite{conte_fully_1997}. An STTMW spectrum $\hat{S}_{AA}(t,\omega)$  is defined on a discrete time-frequency domain:  $[t_1,t_2,...,t_N] \times [{\omega}_1,{\omega}_2,...,\omega_{ K}]$, with time and frequency discretization steps $\Delta t$ and  $\Delta\bar\omega$,\footnote{Note that  $\Delta\bar\omega$ differs form the  discretization step $\Delta\omega$ of the simulated process in Eq.\eqref{sec2_eq2}.} respectively.  Then, we normalize the spectral area at any time $t_n$ as one, i.e., ${\tilde{S}}_{AA}(t_n,\omega_k) = \hat{S}_{AA}(t_n,{\omega}_k)/S_0(t_n)$, where $S_0(t_n) = \sum_{k=1}^{K}\hat{S}_{AA}(t_n,{\omega}_k)\Delta\bar\omega$. 

At time instant $t = t_n$, the parameter vector $\hat{\boldsymbol{\theta}}_{\phi} (t_n)$  is estimated by minimizing the square difference between $\phi(\omega|\boldsymbol\theta_{\phi}(t_n))$ and a smooth version of $\tilde{S}_{AA}(t_n,\omega_k)$ (denoted as $\hat{\phi}(t_n,\omega_k)$) over the frequency ordinate. This optimized fitting can be expressed as:
\begin{equation}
\label{eq_sec3_2}
    \hat{\boldsymbol{\theta}}_{\phi} (t_n) =  \underset{{\boldsymbol{\theta}}_{\phi} (t_n)}{\arg \min} \sum\limits_{k=1}^{K} \left(  \phi(\omega_k|\boldsymbol\theta_{\phi}(t_n))  - \hat{\phi}(t_n,\omega_k) \right)^2.
\end{equation}
In this paper, $\hat{\phi}(\omega_k|t_n)$ is the result of using a 3-second Hann window to smooth $\tilde{S}_{AA}(t,\omega)$ along time. Note that the optimization process takes the implicit normalizing constant $\phi_0(t_n)$ inside $\phi(\omega|\boldsymbol\theta_{\phi}(t_n))$ (see Section \ref{sec2_3}) as a decision variable. After optimization, $\phi_0(t_n)$ is recomputed and constrained by $\sum_{k=1}^{K} \phi(\omega_k|\boldsymbol{\theta}_{\phi}(t_n)) d\omega = 1$ as mentioned in Section \ref{sec2_3_1}.

After fitting the vector of time-varying filter parameters $\hat{\boldsymbol{\theta}}_{\phi} (t_n)$, the second step independently estimates the trend parameters of each scalar element $\hat{{\theta}}_{\phi} (t_n)$ of $\hat{\boldsymbol{\theta}}_{\phi} (t_n)$. Among the three trend functions in Section \ref{sec2_4}, the parameter(s) of the constant and non-parametric trends are computed directly using their definitions. The parameters for linear trends are fitted using the weighted least squares optimization:
\begin{equation}
\label{eq_sec3_3}
    [\hat{{\theta}}_{\phi,t_{mid}}, \hat{{\theta}}'_{\phi,t_{mid}}]=  \underset{[{{\theta}}_{\phi,t_{mid}}, {{\theta}}'_{\phi,t_{mid}}]} {\arg \min} \sum\limits_{l=1}^{L} W(t'_l)  \left[ \hat{\theta}_{\phi}(t'_l) - (\theta_{\phi,t_{mid}} + \theta'_{\phi,t_{mid}}(t'_l-t_{mid})) \right]^2,
\end{equation}
where $t'_1 = t_5$, $t'_L = t_{95}$ and the weighted coefficients are given by:
\begin{equation}
\label{eq_sec3_4}
    W(t') = \frac{ q(t'|\boldsymbol{\theta}_T) } {I_a(t_{95}) - I_a(t_5) },
\end{equation}
where $q(t'|\boldsymbol{\hat\theta}_T)$ is the fitted time-modulating function. The optimization is performed over the interval $[t_{5},t_{95}]$ to improve the accuracy of the fitting at the strong shaking phase of the motion.


\subsection{Parametric fitting to $h_f(t|f_c)$}
\label{sec3_3}
In previous models, $f_c$ has generally been assigned a constant value \citep{conte_fully_1997,rezaeian_stochastic_2008} or estimated using prediction equations based on earthquake magnitude or ground motion duration \citep{vlachos_multi-modal_2016,dabaghi_simulation_2018}. However, the study \cite{su_importance_2024} emphasizes the importance of treating $f_c$ as a free parameter, as it directly influences the long-period linear response spectrum.

The $f_c$ optimization procedure proposed in \cite{su_importance_2024} is summarized and used here. Given a recorded GM, let $S_a^{\mathrm{real}}(T)$ represent its 5\%-damped elastic pseudo-acceleration response spectrum. The corresponding simulated spectrum,  $S_a^{\mathrm{sim}}(T|f_c)$, is computed using synthetic GMs generated from the stochastic GMM with fitted $\hat{\boldsymbol\theta}_T$ and $\hat{\boldsymbol\theta}_F$ but an unknown $f_c$.  $S_a^{\mathrm{sim}}(T|f_c)$ is conditional on $f_c$ and is a random variable due to white noise $\boldsymbol{Z}$. The optimization objective  is to minimize the relative bias between the recorded and simulated response spectra over $T\in[1,10]$s, denoted as:
\begin{equation}
\label{eq_sec3_5}
    \epsilon(f_c) = \left| \int_{1}^{10}   \frac{  \log(S_a^\mathrm{real}(T)) - \mu[\log(S_a^\mathrm{sim}(T|f_c))]}{\sigma[\log(S_a^\mathrm{sim}(T|f_c))]}   d\log(T) \right|.
\end{equation}
Here, $\mu[\cdot]$ and $\sigma[\cdot]$ denote the mean and standard deviation. The logarithmic integral in Eq.\eqref{eq_sec3_5} is practically solved by summing biases on 30 logarithmically spaced points within the integral region. Furthermore, statistical quantities in Eq.\eqref{eq_sec3_5} are estimated through Monte Carlo Simulation by drawing 100 samples from the fitted GMM.  The optimized value of $f_c$ is determined by minimizing Eq.\eqref{eq_sec3_5} within a grid ranging from $[0,2]$ Hz, with an interval of $\Delta f_c = 0.01$ Hz. 


\section{Validation of a Baseline Ground Motion Model}
\noindent
\label{sec4}
This section defines and validates a baseline model that incorporates four modular settings: (1) the spline modulating function in Section \ref{sec2_2}, (2) the uni-modal  second-order frequency filter in Section \ref{sec2_3_1}, (3) a linear trend for the filter frequency $\omega_g(t)$ in Section \ref{sec2_4_2}, and (4) a constant trend for the filter bandwidth $\zeta_g(t)$ in Section \ref{sec2_4_1}. The GMM is fitted following the procedure in Section \ref{sec3}. This baseline model is similar to the spectral-based stochastic GMM \cite{broccardo_spectral-based_2017}, apart from the method for fitting $f_c$.

The baseline model is applied to fit a large recorded dataset (introduced in Section \ref{sec4_1}). The GMM parameters obtained from all the records are then used to simulate a corresponding synthetic dataset. Model validation is performed by comparing the recorded and synthetic datasets in terms of specific ground motion intensity measures (Section \ref{sec4_2_IMs}), linear response spectra (Section \ref{sec4_2_1}), and nonlinear response spectra (Section \ref{sec4_2_2}). To account for the randomness of white noise, all comparative metrics are computed using 30 independent synthetic datasets.

\subsection{Dataset of recorded GMs}
\label{sec4_1}
\begin{figure}[!htb]
   \centering
   \includegraphics[width=0.98\textwidth]{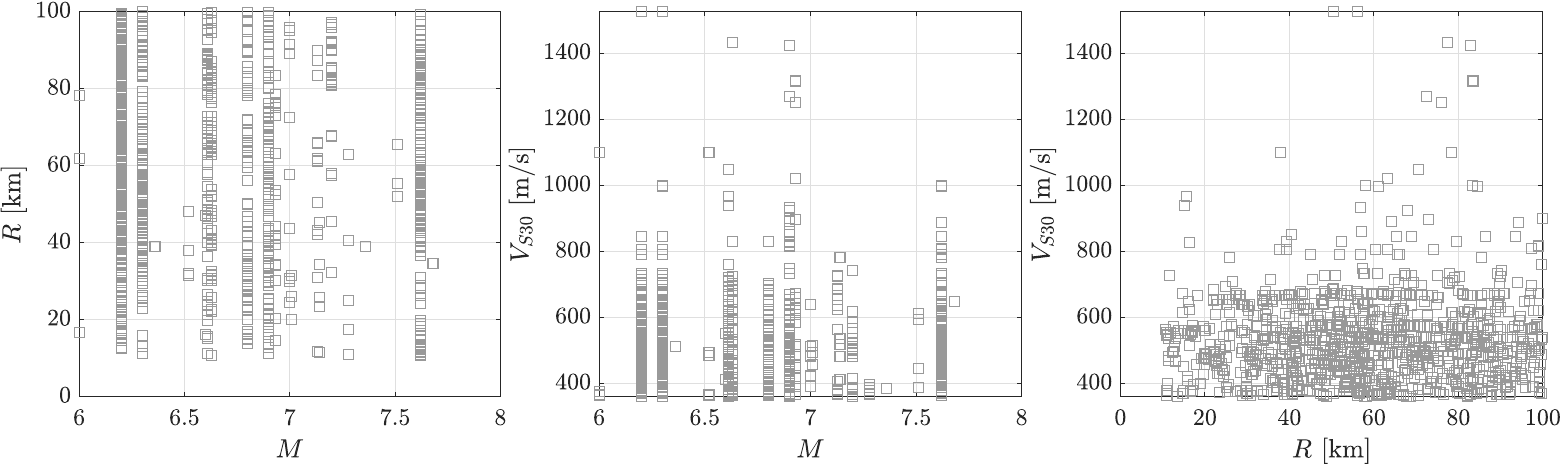}
\caption{Distribution of moment magnitude $M$, source‐to‐site distance $R$, and shear‐wave velocity $V_{S30}$ for the the 1,001 records.}
\label{sec4_MetaData} 
\end{figure}

This study selects a subset from the PEER NGA-West2 database \cite{bozorgnia2014nga,ancheta2014nga} to validate the baseline GMM. The subset comprises 1,001 strong motions, selected based on the following criteria: (1) records from shallow crustal earthquakes in active tectonic regions; (2) earthquakes with moment magnitudes ($M$) greater than 6, targeting events likely to cause structural damage or nonlinear behavior; (3) source-to-site distances ($R$) between 10 and 100 km, where the lower bound reduces near-fault effects like directivity and fling, and the upper bound excludes  low-intensity GMs; (4) GMs recorded at sites with soil shear-wave velocity $V_{S30} > 360$ m/s to reduce the influence of soil nonlinearity; (5) the least usable frequency $f_{\min}$ is less than 0.1 Hz, ensuring that the  records can realistically capture structural responses with periods (at least) up to 10s; and (6) records labeled as pulse-like are excluded. The distributions of $M$, $R$, and $V_{S30}$ for the selected records are shown in Figure \ref{sec4_MetaData}.

Before using the records, they undergo three prepossessing steps. First, the records are rotated into maximally uncorrelated directions. Next, this study focuses on only the component with the larger Arias intensity,  as detailed in \cite{rezaeian_simulation_2012}. Second, the records are down-sampled using an integer decimation factor, ensuring the new sampling rate is as close as possible to 50 Hz (i.e., a sampling interval of 0.02s). Finally, the records are truncated to remove quasi-zero values at the beginning and end, where the initial and final parts are truncated at the thresholds that the cumulative energy reaches 0.01\% and 99.99\% of the total, respectively. The latter two prepossessing steps reduce data size while retaining essential frequency information for the structural response of interest, thus lowering computational costs during GMM fitting.

\subsection{Validation of Intensity Measures}
\label{sec4_2_IMs}
\begin{figure}[htb]
\centering
\includegraphics[width=1\linewidth]{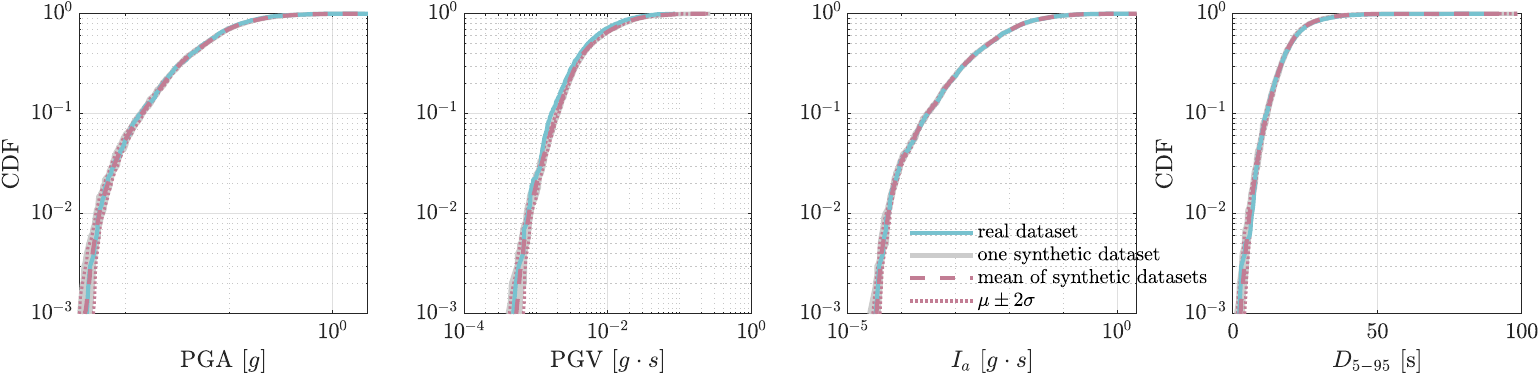}
\caption{Baseline GMM validation w.r.t. four intensity measures of ground motions.}
\label{sec4_fig0} 
\end{figure}

Figure \ref{sec4_fig0} compares the cumulative distribution functions (CDFs) of four IMs: PGA, PGV, Arias Intensity $I_a$, and $D_{5-95}$. Each subplot compares the IMs from the real dataset (blue solid lines) with the 30 synthetic datasets (grey lines), with the red dashed lines representing the averages of the synthetic datasets. Across all IMs, the synthetic dataset averages are closely aligned with the real dataset. Additionally, most real quantiles are captured within the \(\pm 2\sigma\) confidence intervals (red dotted lines). The largest deviations occur in the PGV curves for percentiles between 0.03 and 0.3. Overall, the comparison indicates that the synthetic catalogs effectively replicate the key statistical IMs of the real GM data.

\subsection{Validation of linear-response spectra}
\label{sec4_2_1}
This subsection compares the linear-response spectra $Sa(T)$ within periods $T\in[0.05,10]$s, for spectral damping ratios $\zeta$ of 2\%, 5\%, and 20\%. The comparison metrics include statistical quantities such as the quantiles $q_{n\%}$ of $Sa(T)$ at $n$ percent, the standard deviation $\sigma(\log(Sa(T)))$, and Pearson's correlation coefficients $\rho(T_1,T_2)$ of $\log(Sa(T))$ at distinct periods $T_1$ and $T_2$. 

\begin{figure}[!htb]
   \centering
   \includegraphics[width=0.98\textwidth]{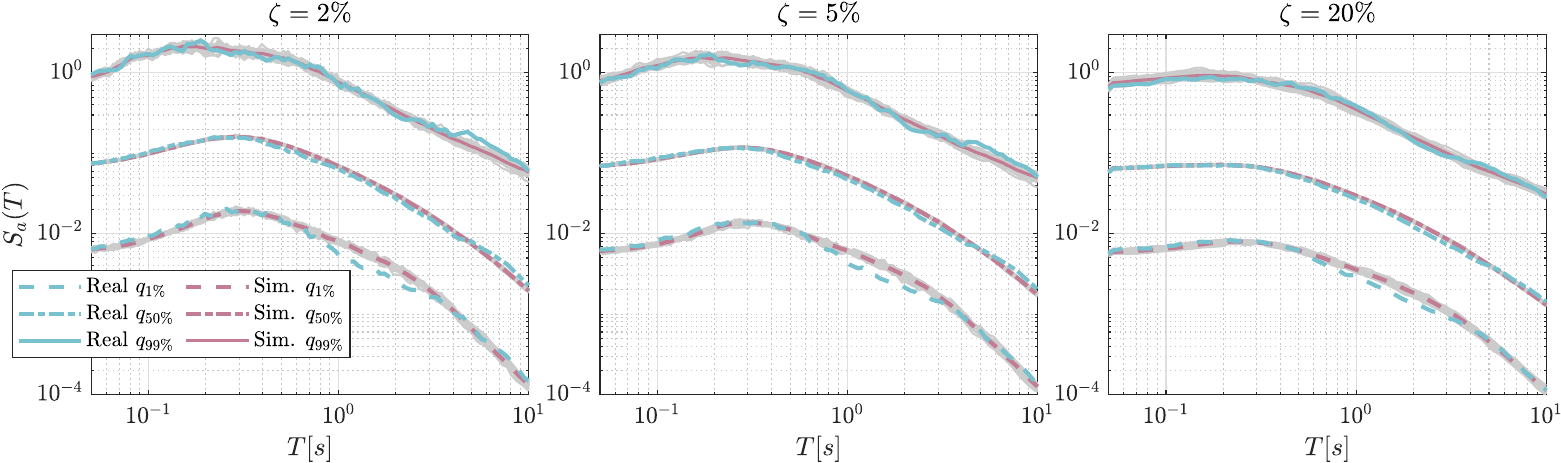}
\caption{Baseline GMM validation w.r.t. three quantiles of $\zeta$-damped linear-response spectra.}
\label{sec4_fig1} 
\end{figure}

Figure \ref{sec4_fig1} compares three quantiles ($q_{1\%}$, $q_{50\%}$, $q_{99\%}$) from the real dataset (blue lines) with that of the 30 synthetic datasets (grey lines). The grey lines showcase the range of possible realizations from the simulations and can be considered as their confidence bounds. The red lines are the averages of the grey lines and thus represent the average level of the synthetic datasets. The figure shows that on average, and regardless of $\zeta$,  the baseline GMM accurately predicts middle-amplitude ($q_{50\%}$) and high-amplitude ($q_{99\%}$) linear spectral responses. Some minor biases are observed for the $q_{99\%}$ curves of $Sa(T\in[3,10]|\zeta=2\%, 5\%)$ and $q_{50\%}$ curves of $Sa(T\in[0.4,10]|\zeta=2\%, 5\%, 20\%)$. However, the baseline model overestimates the low-amplitude ($q_{1\%}$) curves at the middle periods $T\in[0.5,3]$s. Note that this is not a significant concern, as low-amplitude seismic response holds less engineering relevance. Overall, the results validate that the baseline GMM  accurately reproduces most characteristics of the real linear response spectra for different damping levels.

\begin{figure}[!htb]
   \centering
   \includegraphics[width=0.98\textwidth]{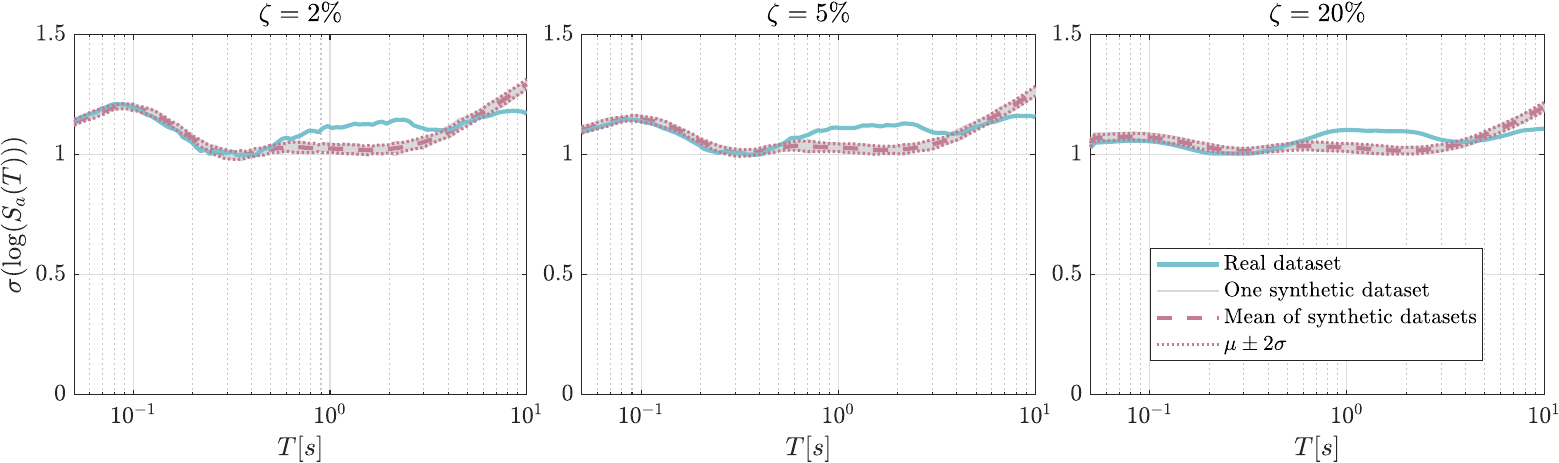}
\caption{Baseline GMM validation w.r.t. logarithmic  standard deviation of $\zeta$-damped linear-response spectra.}
\label{sec4_fig2} 
\end{figure}

Next, the standard deviations $\sigma(\log(Sa(T)))$ are compared, as shown in Figure \ref{sec4_fig2}.  The figure shows that the synthetic datasets precisely predict the spectral variability at short periods $T\in[0.05,0.5]$s when compared to the real GM dataset, regardless of $\zeta$. However, the middle-period ($T\in[0.5,3]$s) variability tends to be slightly underestimated, while the variability at long periods ($T\in[4,10]$s) is slightly overestimated. This is due to the mismatch of the low-amplitude ($q_{1\%}$) curves at the middle periods, as reported earlier. 
%


\begin{figure}[!htb]
   \centering
   \includegraphics[width=0.98\textwidth]{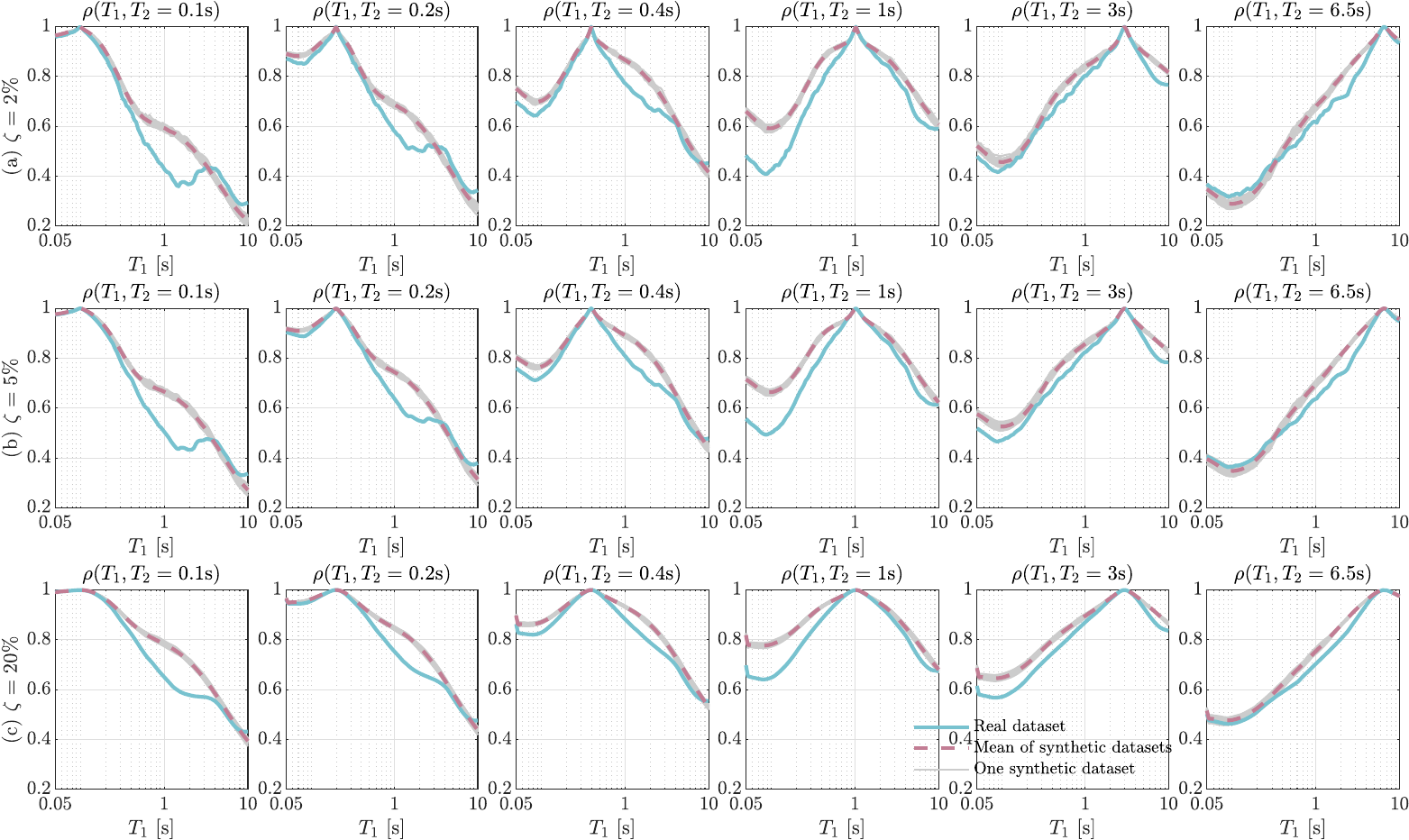}
\caption{Baseline GMM validation w.r.t. linear spectral correlation $\rho(T_1,T_2)$ across different damping ratios $\zeta$.}
\label{sec4_fig3} 
\end{figure}

The subsequent comparison involves the spectral correlation coefficients  $\rho(T_1,T_2)$ between $T_1=[0.05,10]$s and $T_2=0.1,0.2,0.4,1,3,6$s. The results are illustrated in Figure \ref{sec4_fig3}, where each sub-figure compares the mean of real spectral correlation (blue solid lines) to the predicted spectral correlation (red dashed lines). It is observed that, the synthetic datasets closely follow the trend of the real dataset across most period ranges and damping ratios, especially at short periods. Some deviations occur at middle- to long-period correlations, where the synthetic datasets tend to overestimate the correlation compared to the real data. 

In summary, the baseline GMM is validated in terms of the linear-response spectra. It should be noted that, as introduced in Section \ref{sec3_3}, the corner frequency parameter $f_c$ is optimized by matching $Sa(T|\zeta=5\%)$ at periods greater than $1$s. This subsection demonstrated that the baseline GMM also predicts accurately $Sa(T|\zeta)$ with other damping ratios.

\subsection{Validation of nonlinear-response spectra}
\label{sec4_2_2}
This subsection compares the nonlinear response spectra $S_a^{NL}(T)$, more specifically, the inelastic pseudo-acceleration response spectra with different levels of ductility $\mu$. This study focuses on the spectral ordinates $T\in[0.1,10]$s, sets $\zeta$ as $5\%$, and discusses $\mu$ set as 1.5, 2, and 4, respectively.

\begin{figure}[!htb]
   \centering
   \includegraphics[width=0.98\textwidth]{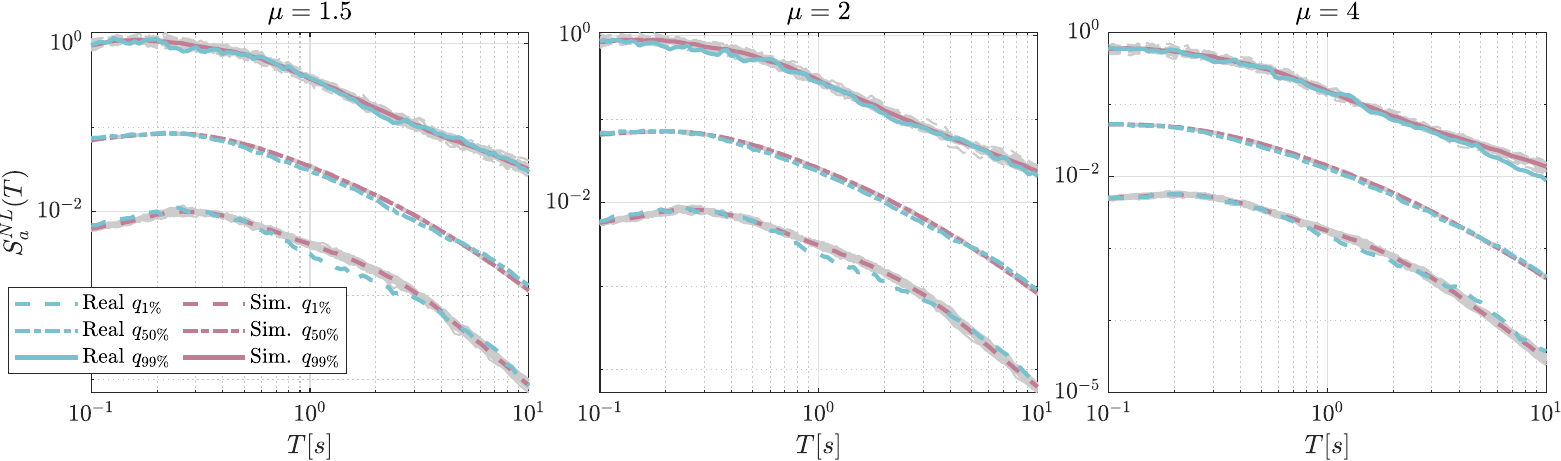}
\caption{Baseline GMM validation w.r.t. three quantiles of nonlinear-response spectra across different given ductility $\mu$.}
\label{sec4_fig4} 
\end{figure}

\begin{figure}[!htb]
   \centering
   \includegraphics[width=0.98\textwidth]{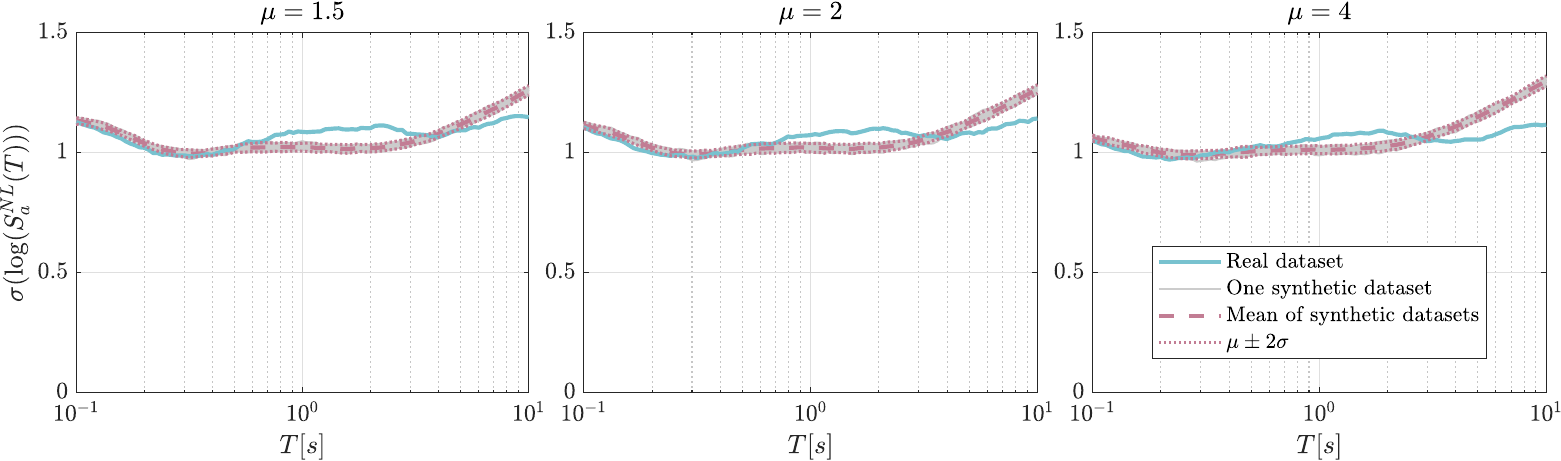}
\caption{Baseline GMM validation w.r.t. logarithmic standard deviation of nonlinear-response spectra  across different given ductility $\mu$.}
\label{sec4_fig5} 
\end{figure}

\begin{figure}[!htb]
   \centering
   \includegraphics[width=0.98\textwidth]{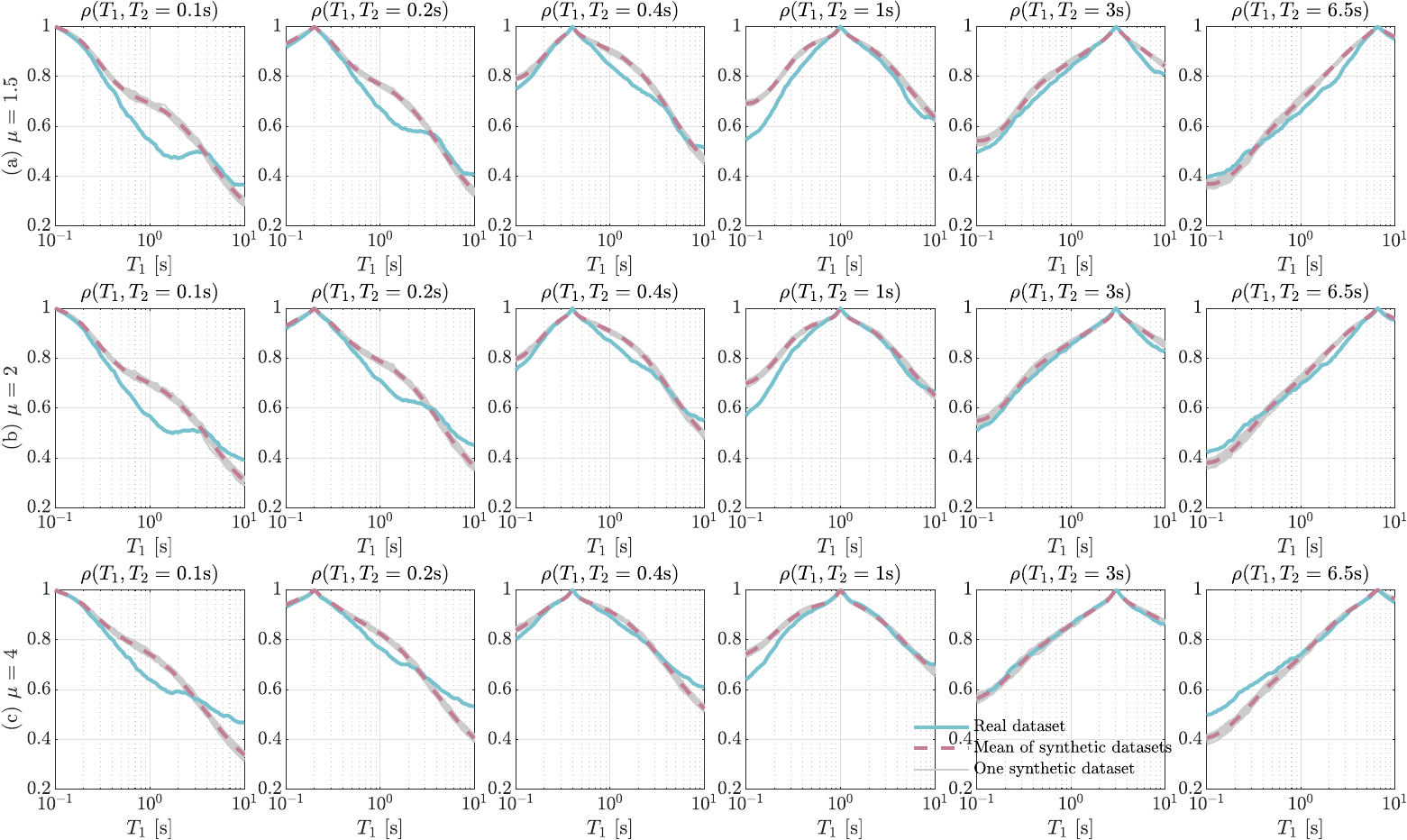}
\caption{Baseline GMM validation w.r.t. nonlinear spectral correlation $\rho(T_1,T_2)$ across different given ductility $\mu$.}
\label{sec4_fig5_corr} 
\end{figure}

The comparison metrics include the same statistical quantities as the linear response spectra. Figure \ref{sec4_fig4} presents the comparison of the three quantities ($q_{1\%}$, $q_{50\%}$, $q_{99\%}$), and Figure \ref{sec4_fig5} compares the standard deviation of $\log(S_a^{NL}(T))$. The comparative results of the nonlinear response spectra are similar to those of  the linear-response spectra. In particular, for most comparative metrics, the mean (over the aleatory uncertainties) of the synthetic GM datasets (red lines) closely matches the real GM dataset (blue lines). In addition, Figure \ref{sec4_fig5_corr}  compares the nonlinear spectra correlation $\rho(T_1,T_2)$, i.e., Pearson's correlation coefficients of  $\log(S_a^{NL}(T))$ at distinct periods. 
 It is observed that the synthetic datasets closely predict  the trend of the real dataset, with a perfect match for curves  where  $T_2=3$s. These figures illustrate the synthetic datasets closely match the real dataset across most period ranges and different $\mu$.

\section{Identification of the Optimal GMM}
\label{sec4_II}
This section intends to find an MFWNM with an ``optimal''  parameterization of the time-varying spectral content. Eight MFWNMs,  incorporating  various  frequency filter and trend functions, are designed in Section \ref{sec4_II_0}.  These models are ranked using newly defined comparison metrics, which average the prediction bias of validation statistics across spectrum ordinates. These comparison metrics are introduced in Section \ref{sec4_II_1} and the ranking results are shown in  Section \ref{sec4_II_2}.

\subsection{Design MFWNMs for comparison}
\label{sec4_II_0}
The previous section shows that the baseline MFWNM closely matches all the given validation metrics and statistics. Besides, the baseline model has a relatively simple configuration with 11 GM parameters. Therefore, we leverage this baseline model as a reference and observe the changes when enriching or modifying its configuration. Specifically, we design several additional GMMs.  The parametric settings for all eight models are listed in Table  \ref{tab1}, where Model 1 refers to the baseline GMM. 

These new GMMs explore the use of  alternative filter types  (introduced in section \ref{sec2_3}), including the KT filter (Model 4), the convex combination of two second-order filers (Models 5 and 6), and the cascade combination of two second-order filers (Models 7 and 8). Additionally, the time-varying filter parameters are characterized by more complex trend functions (introduced in section \ref{sec2_4}). For example, the bandwidth $\zeta_g(t)$ in Model 2 is assigned with the linear model, and both filter parameters in Model 3  adopt the smoothed polyline trend. In total, these GMMs incorporate between 11 and 33 parameters, allowing for a systematic exploration of how different filters and trend functions capture the evolving frequency content of real GMs. 

\begin{table}[!htb]
\caption{A list of the eight considered parametric GMMs. All the models use the spline-modulating function and the high-pass filter integrating with the optimized $f_c$.}
\label{tab1}
\centering
    \centering
    \begin{tblr}{
      colspec={X[0.5]X[4]X[12]X[3]},
      columns = {c,m},
    }
\hline
No. & Filter functions                                      & Trend functions and parameters $\boldsymbol{\theta}_F$                                                                                                                                                                                                                                                                    & Number of total GMM parameters \\ \hline
1   & Second order filter                                   & linear model $\omega_g(t)$ and constant $\zeta_g(t)$ $\boldsymbol{\theta}_F = \{\omega_g(t_{mid}),\omega_g^{'}(t_{mid}),\zeta_g(t_{mid})\}$                                                                                                                                      & 11                        \\ \hline
2   & Second order filter                          & linear models $\omega_g(t)$ and  $\zeta_g(t)$ $\boldsymbol{\theta}_F = \{\omega_g(t_{mid}),\omega_g^{'}(t_{mid}),\zeta_g(t_{mid}),\zeta_g^{'}(t_{mid})\}$                                                                                                                                         & 12                        \\ \hline
3   & Second order filter                          & smoothed polyline models $\omega_g(t)$ and $\zeta_g(t)$ $\boldsymbol{\theta}_F =\{\omega_{g}(t_{p}),\zeta_{g}(t_{p})|p=5,30,45,75,95\}$                                                                                                                                         & 18                        \\ \hline
4   & Kanai-Tajimi filter                              & linear model $\omega_g(t)$ and constant $\zeta_g(t)$ $\boldsymbol{\theta}_F =\{\omega_g(t_{mid}),\omega_g^{'}(t_{mid}),\zeta_g(t_{mid})\}$                                                                                                                                      & 11                        \\ \hline
5   & Convex combination of two second-order filters   & linear models $\omega_{g,j}(t)$, $\zeta_{g,j}(t)$ and $\pi(t)$; $\boldsymbol{\theta}_F =\{\omega_{g,1}(t_{mid}),\omega_{g,1}^{'}(t_{mid}),\zeta_{g,1}(t_{mid}),\zeta^{'}_{g,1}(t_{mid}),\allowbreak   \omega_{g,2}(t_{mid}), \omega_{g,2}^{'}(t_{mid}),\zeta_{g,2}(t_{mid}),\zeta^{'}_{g,2}(t_{mid}),\pi(t_{mid}),\pi^{'}(t_{mid})\}$ & 18                        \\ \hline
6   & Convex combination of two second-order filters        & smoothed polyline models $\omega_{g,j}(t)$, $\zeta_{g,j}(t)$ and $\pi(t)$ $\boldsymbol{\theta}_F =\{\omega_{g,1}(t_{p}),\zeta_{g,1}(t_{p}),\omega_{g,2}(t_{p}),\zeta_{g,2}(t_{p}),\pi(t_{p})|p=5,30,45,75,95\}$                                       & 33                        \\ \hline
7   & Cascade combination of two second-order filters                              &linear models $\omega_{g,j}(t)$ and $\zeta_{g,j}(t)$ $\boldsymbol{\theta}_F =\{\omega_{g,1}(t_{mid}),\omega_{g,1}^{'}(t_{mid}),\zeta_{g,1}(t_{mid}),\zeta^{'}_{g,1}(t_{mid}),\allowbreak \omega_{g,2}(t_{mid}),   \omega_{g,2}^{'}(t_{mid}),\zeta_{g,2}(t_{mid}),\zeta^{'}_{g,2}(t_{mid})\}$                                                                                                                      & 16                        \\ \hline
8   & Cascade combination of two second-order filters & smoothed polyline models $\omega_{g,j}(t)$, $\zeta_{g,j}(t)$ $\boldsymbol{\theta}_F =\{\omega_{g,1}(t_{p}),\zeta_{g,1}(t_{p}),\omega_{g,2}(t_{p}),\zeta_{g,2}(t_{p})|p=5,30,45,75,95\}$                          & 28                        \\ \hline                                                                                                                           
\end{tblr}
\end{table}

\subsection{Comparison metrics}
\label{sec4_II_1}
This subsection introduces comparison metrics to rank the eight parametric GMMs. These metrics measure the averaged absolute biases between the real and synthetic GM datasets when computing certain statistics of the response spectra. We define two groups of metrics: one group is for the statistics of response spectra at a range of periods $T$, and the other group is for the correlation of linear- or nonlinear-response spectra at distinct periods.

For the first group, let  $Q(T)$ and $\hat{Q}(T|\boldsymbol{z})$ denote a statistical quantity of interest (QoI) at periods $T$ that are computed from a real and synthetic GM dataset, respectively. The predicted value of $\hat{Q}(T|\boldsymbol z)$ depends on a set of realized white noises $\boldsymbol{z}$ of a synthetic dataset. Then, the comparison metrics $\epsilon_{QoI}$ is defined as: 
\begin{equation}
\label{eq_QoI_1}
    \epsilon_{QoI} = \frac{1}{N_T N_C} \sum_{i=1}^{N_T}  \sum_{c=1}^{N_C}  \bigg| \frac{Q(T_i)-\hat{Q}(T_i|\boldsymbol{z}_c)}{Q(T_i)}\bigg|,
\end{equation}
where $\{T_i|i=1,...,N_T\}$ are 101 logarithmically spaced points on a spectral range $T\in[T_L,T_U]$s, and $\boldsymbol{z}_c$ is the white noise set for the $c$-th synthetic GM dataset among the total $N_C$  datasets. This metric applies to quantiles $q_{n\%}$ and  standard deviations of both linear- and nonlinear-response spectra.

For the second group, the comparison metric measures the absolute discrepancy:
\begin{equation}
    \epsilon_{QoI} = \frac{1}{N_T^2 N_C} \sum_{i=1}^{N_T}\sum_{m=1}^{N_T}\sum_{c=1}^{N_C}\bigg|  {Q(T_i,T_m)-\hat{Q}(T_i,T_m|\boldsymbol{z}_c)}\bigg|,
\end{equation}
where $Q(T_i,T_m)$ and $\hat{Q}(T_i,T_m|\boldsymbol{z}_c)$ are defined similarly to the first group. This indicator involves a double summation of the biases over period pairs in the region $[T_L,T_U]\times [T_L,T_U]$, and a third summation over the $N_c$ simulated datasets. This metric only evaluates the linear and nonlinear spectral correlation $\rho(T_1,T_2)$.
In this study, the averaged spectral range $[T_L,T_U]$ is set as $[0.05,10]$s for  linear-response spectra and  $[0.1,10]$s for measuring nonlinear response spectra. All the comparison metrics are computed using 30 synthetic GM datasets, i.e., $N_C=30$.
\subsection{Ranking results}
\label{sec4_II_2}
This subsection shows the comparison metrics $\epsilon_{QoI}$ of the eight models for six different response spectra, namely the $\zeta$-damped linear response spectrum $Sa(T)$ for three damping ratios ($\zeta_{h\in[1,2,3]}=2\%,5\%,20\%$)  and the constant ductility inelastic response spectrum $S_a^{NL}(T)$ with $\zeta=5\%$ for three ductility factors ($\mu_{h\in[1,2,3]}=1.5,2,4$). For each response spectrum, we compare the spectral quantiles $q_{n\%}(Sa(T))$ (or $q_{n\%}(S_a^{NL}(T))$) over $n \in [1,2,...,99]$, the spectral variability $\sigma(\log(Sa(T)))$ (or $\sigma(\log(S_a^{NL}(T)))$), and the  spectral correlation $\rho(\log(Sa(T_1)),\log(Sa(T_2)))$ (or $\rho(\log(S_a^{NL}(T_1)),\log(S_a^{NL}(T_2)))$). 

\begin{figure}[!htb]
   \centering
   \includegraphics[width=0.98\textwidth]{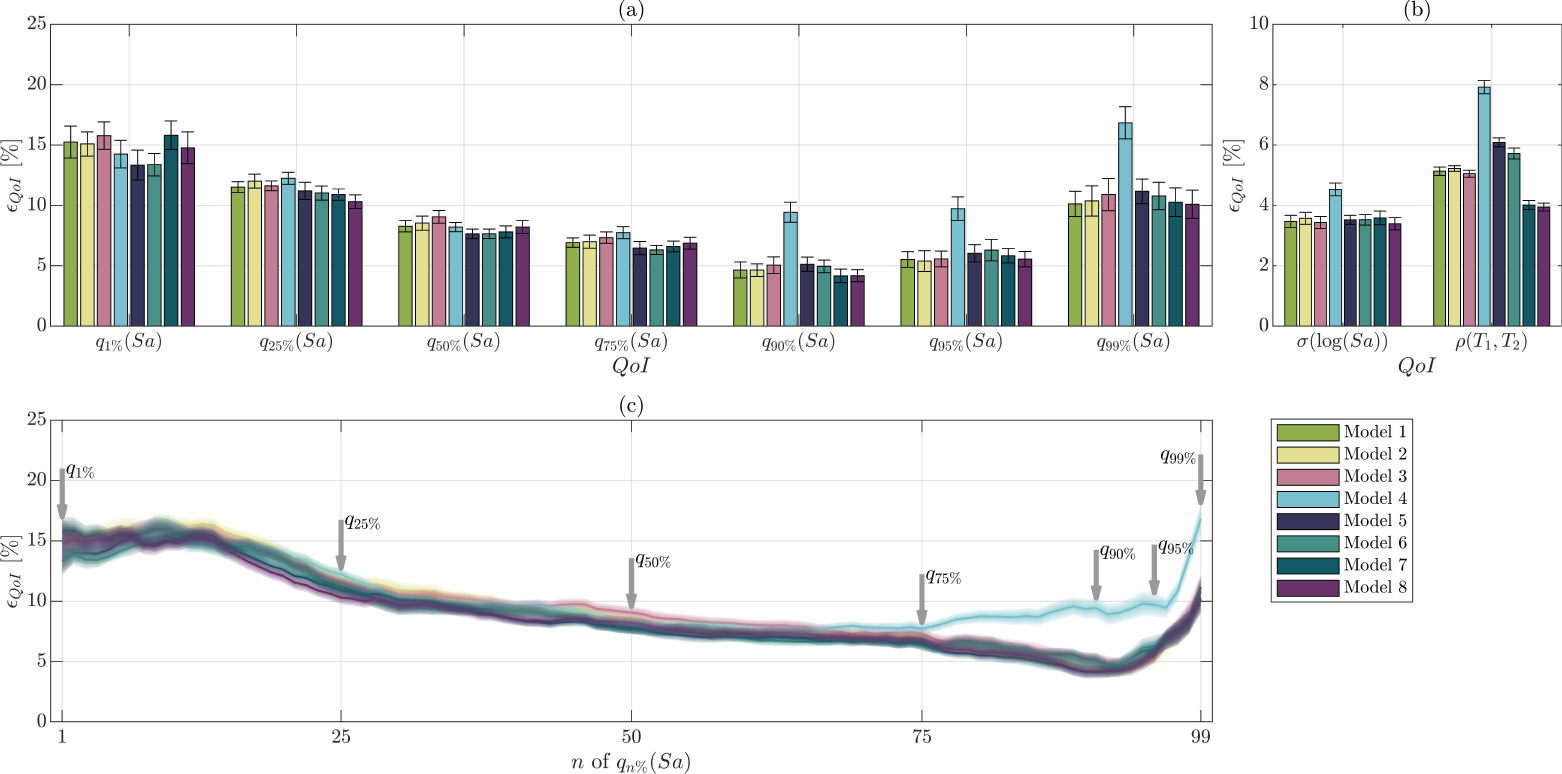}
\caption{Compare error metrics $\epsilon_{QoI}$ of 5\%-damped $Sa(T)$ for the eight considered parametric GMMs. (a) and (b): Error bars of different metrics $\epsilon_{QoI}$, with the whiskers representing $\pm 1\sigma$ confidence bounds; (c): Biases of $q_{n\%}$ vary with the quantile level $n$,  with $\pm 1\sigma$ confidence bounds.}
\label{sec4_fig7} 
\end{figure}

To improve readability, we  show only the case of $Sa(T|\zeta_2=5\%)$ in Figure \ref{sec4_fig7} and the case of $S_a^{NL}(T|\mu_2=2)$ in Figure \ref{sec4_fig10}, with figures for the remaining spectra included  in   \ref{appendix_2}. In each figure,  sub-figure (c)  illustrates how the spectral quantile $q_{n\%}$ varies with the quantile level $n$, with specific quantile levels $n\in[1,25,50,75,90,99]$ compared in sub-figure (a). Sub-figure (b) displays the spectral variability and spectral correlation. The same observations from the six response spectra are:  
(1) tail quantiles (e.g., $q_{1\%}$ and $q_{99\%}$) exhibit larger biases than the central quantiles (i.e, $q_{n\%}$ with $25 \le n \le 75$); 
(2) biases in low-amplitude quantiles $q_{n\%}$ with $n\le25$ are likely larger than corresponding high-amplitude quantiles  $q_{(100-n)\%}$;
(3) Model 4 tends to present the largest biases in spectral variability and spectral correlation; 
(4) with the exception of Model 4, the other models tend to present indistinguishable performance, especially for $Sa(T|\zeta=2\%, 5\%)$. Compared to other seven models, Model 4 exhibits significant variability in predicting $q_{n\%}$, as it performs the worst for $Sa(T|\zeta=2\%)$ and the best for $S_a^{NL}(T|\mu=4)$. As for the other four response spectra (i.e., $Sa(T|\zeta=5\%,20\%)$ and $S_a^{NL}(T|\mu=1.5,2)$), Model 4 performs well at low-to-middle-amplitude quantiles (with $n\le 75$) but poorly at high-amplitude quantiles.  In summary, no single model consistently outperforms the others.

\begin{figure}[!htb]
   \centering
   \includegraphics[width=0.98\textwidth]{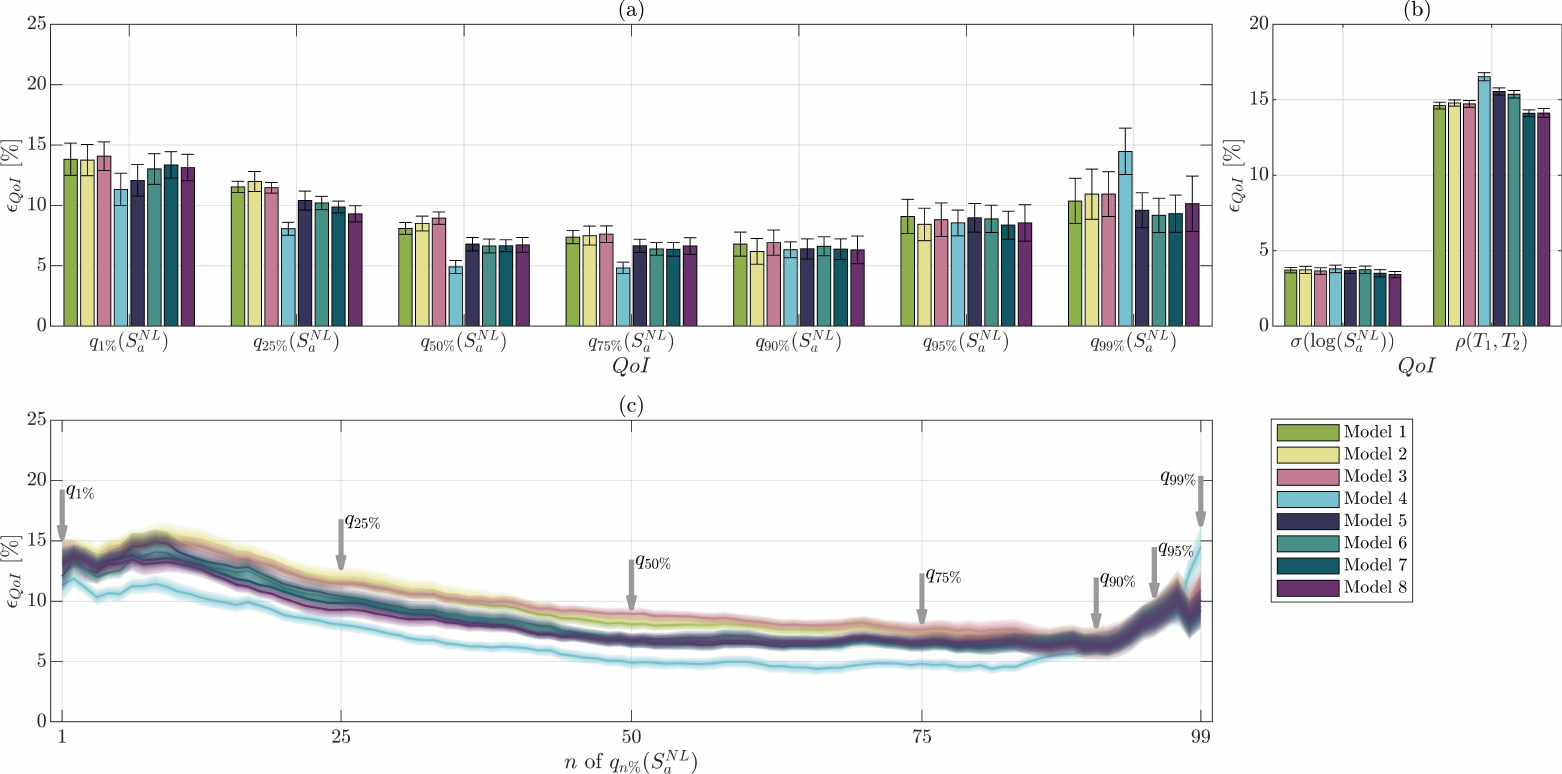}
\caption{Compare error metrics $\epsilon_{QoI}$  of constant-ductility ($\mu=2$) spectra for the eight considered parametric GMMs. (a) and (b): Error bars of different metrics $\epsilon_{QoI}$, with the whiskers representing $\pm 1\sigma$ confidence bounds; (c): Biases of $q_{n\%}$ vary with the quantile level $n$,  with $\pm 1\sigma$ confidence bounds.}
\label{sec4_fig10} 
\end{figure}

\begin{figure}[htb]
\centering
\begin{subfigure}[b]{0.49\textwidth}
   \includegraphics[width=1\linewidth]{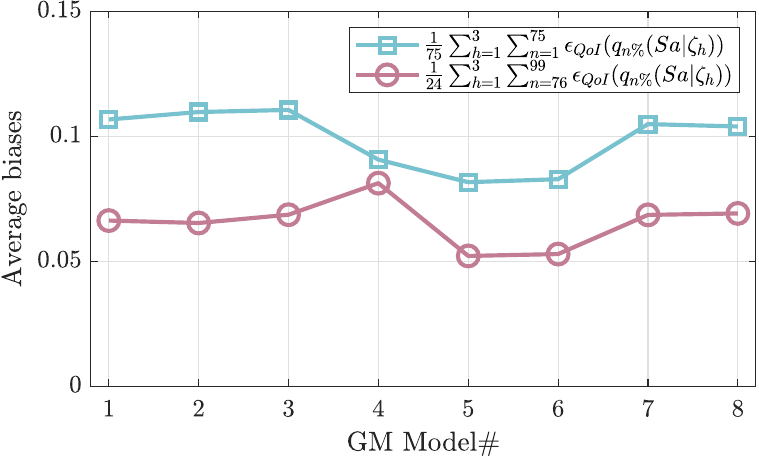}
   \caption{Average biases of $Sa$}
   \label{sec4_fig14a} 
\end{subfigure}
 \hfill
\begin{subfigure}[b]{0.49\textwidth}
   \includegraphics[width=1\linewidth]{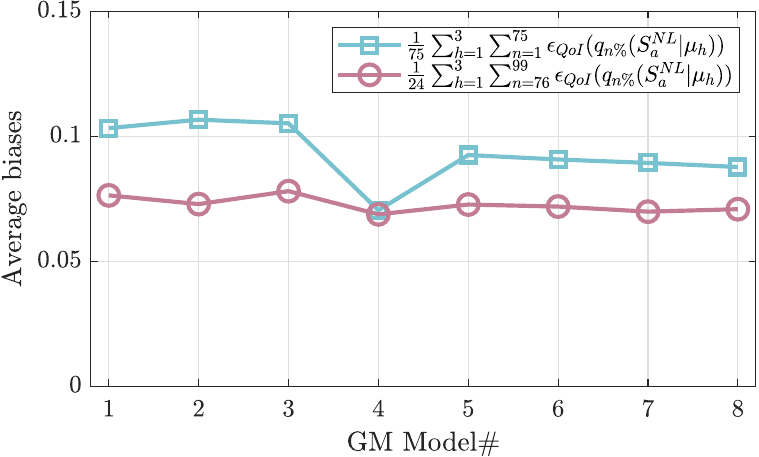}
   \caption{Average biases of $S_a^{NL}$}
   \label{sec4_fig14b} 
\end{subfigure}
\caption{Average biases of the eight GMMs regarding predicting (a) linear-response spectrum $Sa$, and (b) Nonlinear-response spectrum $S_a^{NL}$. The  biases are averaged over the high-amplitude quantiles (Red lines) and low-to-middle quantiles (Blue lines), respectively.}
\label{sec4_fig14} 
\end{figure}

Next, we compare the average performance of the eight GMMs over high-amplitude quantiles  (i.e., $q_{n\%}$ for $n > 75$) and low-to-middle-amplitude quantiles (i.e., $q_{n\%}$ for $n \le 75$), respectively. This separate comparison aims to highlight the high-amplitude quantiles, which refer to extreme seismic responses and, therefore, are central for engineering applications. In particular, for the linear-response spectra $S_a(T|\zeta_h)$, the average model performance is calculated as the average of a given set of $\epsilon_{QoI}$ (from Eq. \eqref{eq_QoI_1}). Specifically, we define two sets: the first set represents the high-amplitude quantiles $\{q_{n\%}(Sa|\zeta_h) | h\in [1,2,3], n\in [76,77,...,99]\}$, while the second set represents the low-to-middle-amplitude quantiles $\{q_{n\%}(Sa|\zeta_h) | h\in [1,2,3], n\in [1,2,...,75]\}$.
As for the nonlinear-response spectra $S_a^{NL}(T|\mu_h)$, the  average model performance is similarly defined by averaging different $\epsilon_{QoI}$ with various $\mu_h \in [1.5,2,4]$ and spectral quantile levels.  These average performance are presented in Figure \ref{sec4_fig14}, which shows the average  biases of the eight GMMs in terms of predicting the linear-response spectra (Figure \ref{sec4_fig14a}) and the nonlinear-response spectra (Figure \ref{sec4_fig14b}).  

Taking Model 1 as a reference, the key observations in Figure \ref{sec4_fig14} are: 
(1) Complex trend functions (referring to Models 2 and 3) offer minimal or even negative improvement; 
(2) Model 4 (owning the same number of parameters as Model 1) performs better on the low-to-middle-amplitude quantiles, regardless of $Sa$ or $S_a^{NL}$. However, Model 4 performs worse on the high-amplitude quantiles of $Sa$ and slightly better on the high-amplitude quantiles of $S_a^{NL}$. (3) Minimal improvement is observed when using multi-mode frequency-filter functions and more complex trend functions (referring to Models 5$\sim$8). In summary, Model 1 has the simplest parametric form and balanced performance, specifically in predicting high-amplitude quantiles. Therefore, we define Model 1 as owning the ``best" configuration and select it to assemble the hierarchical GMM in the subsequent sections. 



\section{Characterization and Modeling of Epistemic Uncertainties in GMM parameters}
\noindent
\label{sec5}
This section models the GMM parameter $\boldsymbol\theta\in\mathbb{R}^M$ as a random vector $\boldsymbol{\Theta}$ with a parametric probabilistic model, defined by its joint cumulative distribution function (CDF) $F_{\boldsymbol{\Theta}}(\boldsymbol\theta)$  with marginals $F_{\Theta_1}(\theta_1),...,F_{\Theta_M}(\theta_M)$. According to Sklar’s theorem \cite{sklar_fonctions_1959},  if all $F_{\Theta_m}(\theta_m)$ are continuous, there uniquely exists a copula function $C(\cdot): [0,1]^M \rightarrow [0,1]$, denoted as:
\begin{equation}
   \label{eq_sec5_1}
    F_{\boldsymbol{\Theta}}(\boldsymbol\theta) = C(F_{\Theta_1}(\theta_1),...,F_{\Theta_M}(\theta_M)),
\end{equation}
which maps the marginal CDFs to the joint CDF. The copula function  $C(\cdot)$ is independent of the marginals and purely describes the dependence structure of the random vector.  The joint probability density function (PDF) is given by:
\begin{equation}
    f_{\boldsymbol{\Theta}}(\boldsymbol\theta) = c(u_1,...,u_M) \prod_{m=1}^{M}f_{\Theta_m}(\theta_m),
\end{equation}
where $\boldsymbol{u} = [u_m=F_{\Theta_m}(\theta_m)|m=1,...,M]\in [0, 1]^M$, $c(\cdot)$ is the copula density function defined as  $c(\boldsymbol{u}) = \frac{\partial^M C(\boldsymbol{u})}{\partial{u_1}...\partial{u_M}}$, and $f_{\Theta_m}(\theta_m)$ are the marginal PDFs. Based on Sklar’s theorem, this study separately models the marginals and the dependence of the random GM parameters to obtain their joint PDF, as illustrated in the following subsections.

\subsection{Marginal models of random GMM parameters}
\label{sec5_1}
\begin{figure}[!htb]
   \centering
   \includegraphics[width=0.98\textwidth]{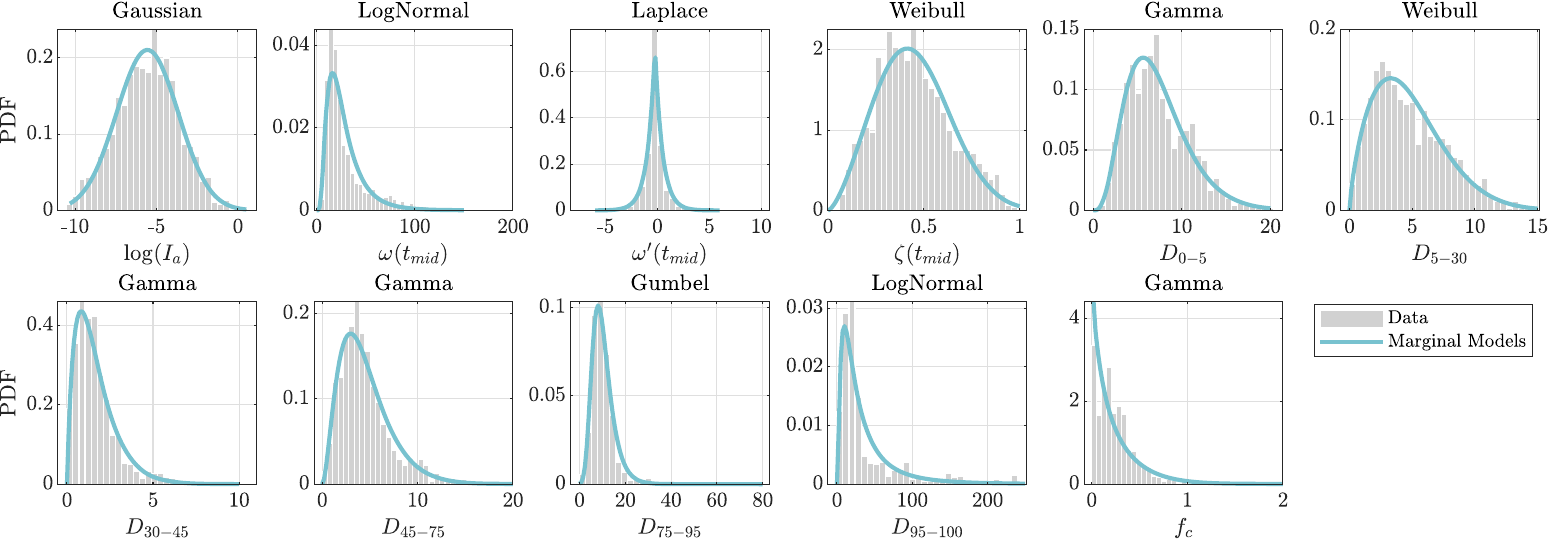}
\caption{Histograms of  the baseline GMM parameters extracted from the 1,001-GM dataset and their corresponding fitted marginal models.}
\label{sec5_fig1} 
\end{figure}

The baseline GMM (Model 1) is fitted to the 1,001-GM dataset, resulting in a corresponding dataset of  GM parameters with size $1,001 \times 11$. Each parameter $\theta_m$ is then fitted with a marginal distribution. Specifically, the marginal distributions are obtained in two steps: (1) model selection using Bayesian information criterion (BIC) \footnote{In this study, the candidate distributions include Gaussian, Lognormal, Gumbel, Weibull, Gamma, Exponential, Beta, Logistic, Laplace, Rayleigh.}; (2)  parameters estimation for the selected marginals using Maximum Posteriori Estimation (MAP) with uniform prior (equivalent to maximum likelihood estimation, MLE).
Table \ref{tab2} summarizes the fitted marginal models along with their corresponding distribution parameters, the distribution support, and the first two statistical moments.  The distribution supports are determined by visual observations of corresponding histograms. Moreover, the supports are adjusted to avoid: (1) numerical issues (e.g.,  $t_f>200 \rm{s}$ causes out-of-memory) and (2) unrealistic simulation (e.g., $t_f$ is too small/large or  $\omega_g(t)=0$).  The fitted marginals along with their histograms are visualized in  Figure \ref{sec5_fig1}.

\begin{table}[h]
\begin{center}
 \caption{Fitted marginal models of random GM Parameters}
 \label{tab2}
\begin{tabular}{ccccccccc}
\hline
\multirow{2}{*}{\textbf{No.}} & \multirow{2}{*}{\textbf{\begin{tabular}[c]{@{}c@{}}GM \\ Parameter\end{tabular}}} & \multirow{2}{*}{\textbf{Marginal}} & \multicolumn{2}{c}{\textbf{Parameters}} & \multicolumn{2}{c}{\textbf{Moments}}& \multirow{2}{*}{\textbf{\begin{tabular}[c]{@{}c@{}}Distribution\\ support\end{tabular}}} & \multirow{2}{*}{\textbf{Unit}} \\
                              &                                                                                   &                                    & Par\#1& Par\#2 & $\mu$&$\sigma$&                                                                                        &                                \\ \hline
1                             & $\log(I_a)$& Gaussian& -5.557& 1.896
& -5.557&1.896
& {(}-Inf, Inf{)}& $\log(g^2\cdot \rm{s} )$\\
2                             & $\omega(t_{mid})$                                                                 & Lognormal& 3.162& 0.610
&  
 
 28.446&19.112
& {(}$0$, Inf{)}& $2\pi \cdot \mathrm{Hz}$                \\
3                             & $\omega^{\prime}(t_{mid})$                                                        & Laplace& -0.227& 0.709
&  -0.227&1.003
& {(}-Inf, Inf{)}& $2\pi \cdot \mathrm{Hz/s}$              \\
4                             & $\zeta(t_{mid})$                                                                  & Weibull& 0.505& 2.524
&  
 
 0.448&0.190
& {[}0.02, 1{]}& 1                              \\
5                             & $D_{0-5}$                                                                         & Gamma& 0.595& 4.357
&  7.320&3.507
& {[}0.1, 20{]}& s                              \\
6                             & $D_{5-30}$                                                                        & Weibull& 5.398& 1.729
&  
 
 4.811&2.869
& {[}0.1, 15{]}& s                              \\
7                             & $D_{30-45}$                                                                       & Gamma& 1.167& 1.965
&  1.684&1.202
& {[}0.1, 10{]}& s                              \\
8                             & $D_{45-75}$                                                                       & Gamma& 0.637& 2.899
&  
 
 4.549&2.672
& {[}0.1, 20{]}& s                              \\
9                             & $D_{75-95}$                                                                       & Gumbel& 8.172& 3.637
&  10.271&4.664
& {[}0.1, 40{]}& s                              \\
10                            & $D_{95-100}$                                                                      & Lognormal& 3.196& 0.960
&  
 
 38.737&47.634
& {[}0.1, 40{]}& s                              \\
11                            & $f_c$                                                                             & Gamma& 3.572& 
0.853
&  0.239&0.259
& {[}0, 2{]}                                                                             & Hz                             \\ \hline
\end{tabular}
\end{center}
\end{table}

\subsection{Dependence models of random GMM parameters}
\label{sec5_2}
This subsection introduces two copula density functions: Gaussian copula (equivalent to the Nataf distribution \cite{liu1986multivariate}) and R-vine copula \cite{bedford_vinesnew_2002}.


\subsubsection{Gaussian copula}
The Gaussian copula is widely used to model the dependence of random vectors with known marginals.  The Gaussian copula density function is defined as \cite{nelsen_introduction_2006,liu1986multivariate}:
\begin{equation}
  c(\boldsymbol{u}) =  \frac{1}{\sqrt{\det \boldsymbol R}} \exp \left( -\frac{1}{2} \left(\begin{matrix}
       \Phi^{-1}(u_1) \\
           \vdots     \\
       \Phi^{-1}(u_M) \\
  \end{matrix}\right)^\mathrm{T} \cdot (\boldsymbol{R}^{-1}-\bf I) \cdot \left(\begin{matrix}
      \Phi^{-1}(u_1) \\
           \vdots     \\
       \Phi^{-1}(u_M) \\
  \end{matrix}\right) 
  \right)
\end{equation}
where $\Phi^{-1}(\cdot)$ is the inverse CDF of a standard normal distribution,  $\bf I$ is an identity matrix, and $\boldsymbol{R}\in [-1,1]^{M\times M}$ is a symmetric correlation matrix.  The correlation coefficients $[\boldsymbol{R}]_{i,j} = [\boldsymbol{R}]_{j,i}$ control the pairwise correlations between $Z_i = \Phi^{-1}(F_{\Theta_i}(\theta_i)) $ and $Z_j = \Phi^{-1}(F_{\Theta_j}(\theta_i))$ in the standard Gaussian space.  
The fitted Gaussian copula and the  $[\boldsymbol{R}]_{i,j}$ are shown in the lower triangle of Figure \ref{sec5_fig2}.


\subsubsection{R-vine copula}
\label{sec5_2_3}
The construction of $M$-variate copulas becomes challenging as the dimension $M$ increases. To simplify this process, Bedford and Cooke \cite{bedford_vinesnew_2002} introduced the regular vine (R-vine), which is a graphical model consisting of $M-1$ trees, denoted as  $\{\mathcal{T}_m|m=1,2,...,M-1\}$.
The $m$-th tree $\mathcal{T}_m$ contains $m-2$ edges, each representing a bivariate copula \cite{nelsen_introduction_2006}. As a result, an R-vine copula involves a total of  $M(M-1)/2$ bivariate copulas.
R-vine copulas can model more complex correlations, including the non-symmetric and tail dependence. Specifically, the R-vine model decomposes an $M$-variate copula density into the product of $M(M-1)/2$ bivariate copulas densities: 
\begin{equation}
    c(\boldsymbol{u}) = \prod_{e=1}^{M(M-1)/2} c_{i_e,j_e|D_e}(u_{i_e|D_e},u_{j_e|D_e}).
\end{equation}
Here, $c_{i_e,j_e|D_e}(\cdot)$ depends on a subset $D_e\subset \mathcal{\bar{A}}$ and serves as an edge relating two distinct nodes $i_e,j_e\in \mathcal{A}$, where $\mathcal{A}$ and $\mathcal{\bar{A}}$ are complementary sets, i..e, $\mathcal{A} \cup \mathcal{\bar{A}} = \{ 1,2,...,M\}$ and $\mathcal{A} \cap \mathcal{\bar{A}} = \varnothing$.   $c_{i_e,j_e|D_e}(\cdot)$ links two uni-variate conditional CDFs and measures their dependence. For example, when $i_e=1$, $j_e=2$ and $D_e=\{3,4\}$, the two inputs of $c_{1,2|\{3,4\}}(\cdot)$ are CDFs of two conditional random variables: $u_{1|\{3,4\}} = F_{\theta_1|\theta_3,\theta_4}(\theta_1|\theta_3,\theta_4)$ and $u_{2|\{3,4\}} = F_{\theta_2|\theta_3,\theta_4}(\theta_2|\theta_3,\theta_4)$.


The R-vine decomposition is not unique (with $2^{\binom{M-2}{2}-1}M!$ choices) but not arbitrary (see the constraints in \cite{bedford_vinesnew_2002}). There are two special types of R-vine, including the drawable vine (D-vine) \cite{kurowicka_distribution-free_2005} and canonical vine (C-vine) \cite{aas_pair-copula_2009}. In the D-vine, each node in one tree is connected with two edges at most, while in the C-vine, each tree has a dominating node connecting to all the other nodes in the tree. These two special vine structures mitigate the optimization challenge, requiring only sorting the  nodes in $\mathcal{T}_1$ (within $M!$ choices).

We fit the copula model to the extracted GM parameters using the UQLab toolbox \cite{torre_general_2019,torre_uqlab_2022}.
The fitting involves four steps : (1) select a vine structure, focusing on C-vine and D-vine in this study; (2) sort the nodes to maximize the sum of Kendall's rank correlation coefficients for the $M-1$ bivariate copulas in $\mathcal{T}_1$; (3) use BIC to select the best bivariate copula from six candidates (Independence-, Gaussian-, Gumbel-, Clayton-, Frank-, and $t$-bivariate  copulas, as introduced in \cite{nelsen_introduction_2006}); and (4) independently estimate the bivariate copula parameters using MLE. This approach is applied to both C-vine and D-vine, with C-vine achieving a higher likelihood score.

\begin{figure}[!htb]
   \centering
   \includegraphics[width=0.99\textwidth]{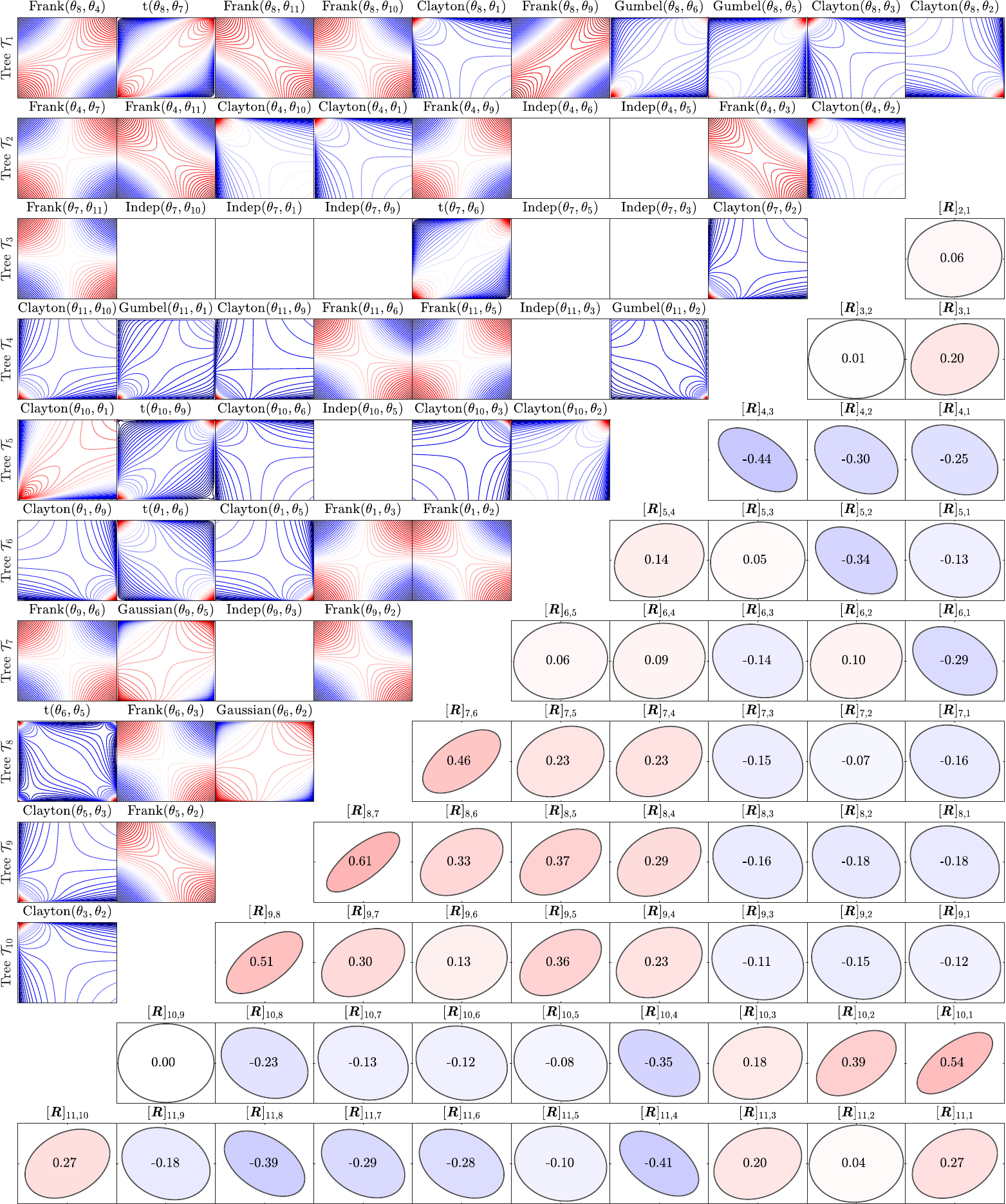}
\caption{Gaussian copula (lower triangle) and Vine Copula (upper triangle)  fitted to the parameters extracted from the 1,001-GM dataset. The Gaussian copula displays the  correlation coefficients, while  the Vine Copula plots the contours of the logarithmic copula density (i.e., $\log_{10}(c_{i_e,j_e|D_e}(u_{i_e|D_e},u_{j_e|D_e}))$. Independence is abbreviated as ``Indep''.}
\label{sec5_fig2} 
\end{figure}

The fitted C-vine model is shown in the upper triangle of  Figure \ref{sec5_fig2}.  A sequence of the sorted nodes is $S=[8,4	,7,	11,	10,	1,	9,	6,	5]$, where the $m$-th element  $s_m$ of $S$ represents the dominating node of $\mathcal{T}_{m+1}$. The first tree $\mathcal{T}_1$ contains 10 fitted bivariate copulas, where each copula links two unconditional random variables. For example, ``Frank($\theta_8$, $\theta_4$)'' refers to a Frank bivariate copula $C_{8,4}(u_8,u_4)$ measuring the dependency between the random variables $\theta_8$ and $\theta_4$. 
  In subsequent trees, the bivariate copulas $C_{i_e,j_e|D_e}(\cdot)$ in  $\mathcal{T}_m$  ($m\geq 2$) are conditional on the first  $(m-1)$ elements of $S$, i.e., $D_e= \{s_1,s_2,...,s_{m-1}\}$.  For example, ``Clayton($\theta_{11}$, $\theta_{10}$)'' in $\mathcal{T}_4$  represents a Clayton copula $C_{11,10|D_e=\{8,4,7\}}(\cdot)$ conditional on three variables $\theta_8$, $\theta_4$ and $\theta_7$. Specifically, this copula links two conditional CDFs $F(\theta_{11}|\theta_8,\theta_4,\theta_7)$ and $F(\theta_{10}|\theta_8,\theta_4,\theta_7)$, expressed as $C_{11,10|8,4,7}(F(\theta_{11}|\theta_8,\theta_4,\theta_7),F(\theta_{10}|\theta_8,\theta_4,\theta_7))$. These two conditional CDFs can be derived from the previous trees, such as $F(\theta_{11}|\theta_8,\theta_4,\theta_7) = \frac{\partial C_{7,11}(u_7,u_{11}|u_8,u_4)}{\partial u_7} $, where $C_{7,11}(\cdot)$ refers to the copula ``Frank($\theta_7$, $\theta_{11}$)'' in $\mathcal{T}_3$. Therefore, the  copula $C_{11,10|8,4,7}(\cdot)$ in $\mathcal{T}_4$, nested with the two copulas (i.e., Frank($\theta_7$, $\theta_{11}$) and Indep($\theta_7$, $\theta_{10}$)) in $\mathcal{T}_3$, characterizes the dependence between $\theta_{11}$ and  $\theta_{10}$ conditioned on $\theta_8$, $\theta_4$ and $\theta_7$. This nested structure applies to each bivariate copula $C_{i_e,j_e|D_e}(\cdot)$ in  $\mathcal{T}_m$  ($m\geq 2$). The product of all bivariate copulas across the trees fully captures the dependence structure of the random GMM parameters $\boldsymbol{\Theta}$.

\section{Validation of Hierarchical GMMs}
\noindent
\label{sec6}
After characterizing the uncertainties in the GMM parameters $\boldsymbol{\theta}$, the stochastic simulator in Eq.\eqref{sec2_eq2+++} is updated into a  hierarchical GMM:
\begin{equation}
\label{}
    \boldsymbol{A}_g(t) = \mathcal{M}(t|\boldsymbol{\Theta},\boldsymbol{Z}), t \in [0,t_f].
\end{equation}
The hierarchical GMM involves two levels of uncertainties: $\boldsymbol{Z}$ and $\boldsymbol{\Theta}$. The white noise $\boldsymbol{Z}$ accounts for the intrinsic uncertainty of individual GMs,  while the random GM parameters $\boldsymbol{\Theta}$ capture record-to-record variability.

The hierarchical GMM contains two simulation steps: (1) simulate a GM parameter sample $\boldsymbol{\theta_n}$ from the fitted joint PDF $ f_{\boldsymbol{\Theta}}(\boldsymbol\theta)$, and (2) synthesize a GM time series by plugging  $\boldsymbol{\theta_n}$  and one white-noise sample $\boldsymbol{{z}_n}$ into $\mathcal{M}(\cdot)$. Notice that, the original dataset (with 1,001 GMs) can be viewed as a set of realized samples from the hierarchical GMM. 

This section examines two versions of the hierarchical GMM to assess the impact of different dependence models for describing the random vector $\boldsymbol{\Theta}$. We focus our analysis on the baseline model from Section \ref{sec4}, while the joint PDFs $f_{\boldsymbol{\Theta}}(\boldsymbol\theta)$ combine the fitted marginals from Section \ref{sec5_1} with the two copula models from Section \ref{sec5_2}, namely the Gaussian and C-vine copulas. To validate these hierarchical GMMs, synthetic datasets of 1,001 GMs are simulated and compared with the recorded dataset of the same size. The comparison metrics include statistics of specific intensity measures (Section \ref{sec6_3}) as well as linear- and nonlinear-response spectra (Sections \ref{sec6_1} and \ref{sec6_2}). For statistical convergence, each version of the hierarchical model generates 30 synthetic datasets, each containing 1,001 independent synthetic GMs. Finally, to assess model generalization, the entire dataset is split into a training and test set, presented in  Sections \ref{sec6_4}.

\subsection{Validation of intensity measures}
\label{sec6_3}
\begin{figure}[!htb]
   \centering
   \includegraphics[width=0.98\textwidth]{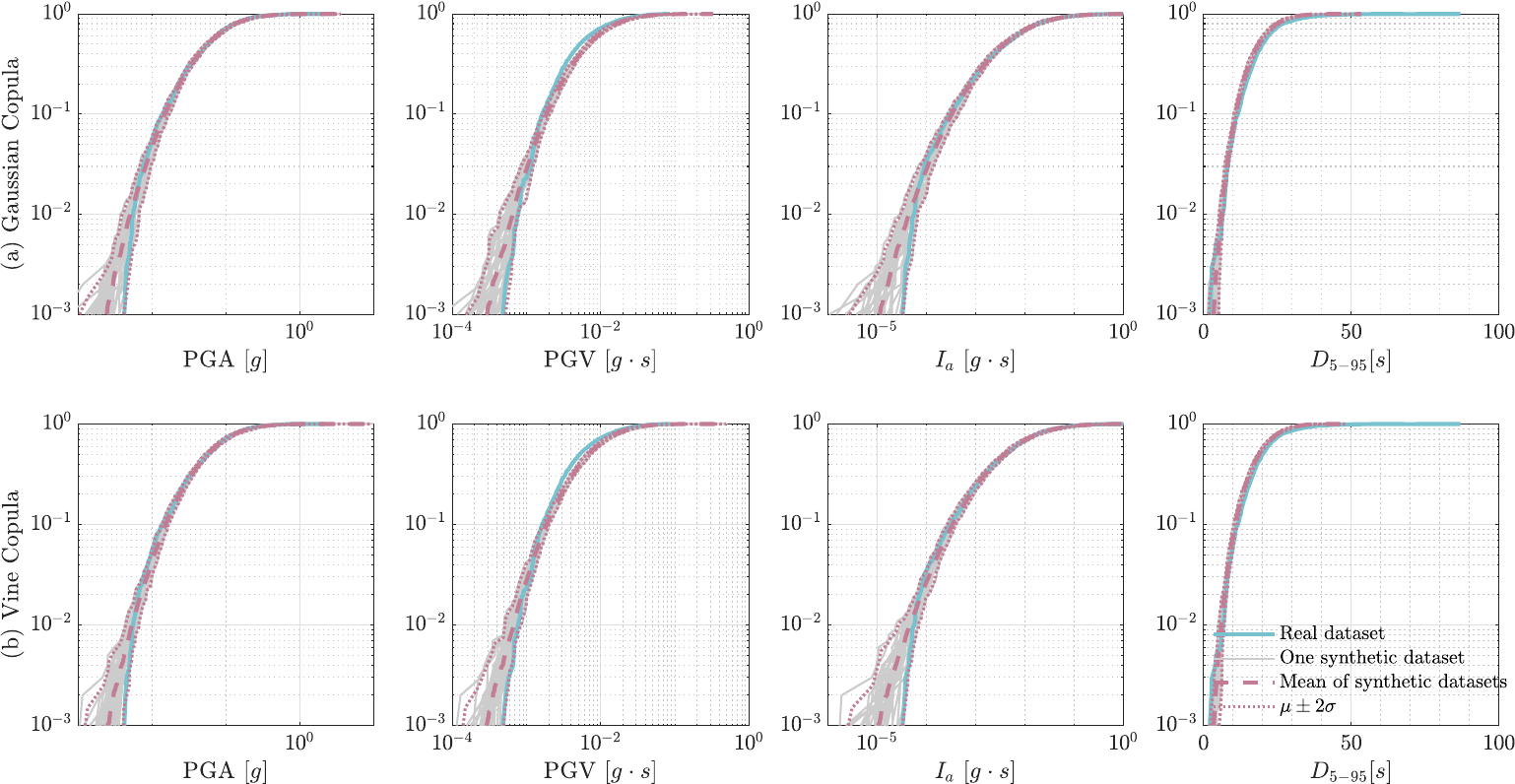}
\caption{Validation of the hierarchical GMM using two different copula models w.r.t. CDFs of four intensity measures: PGA, PGV, Arias intensity $I_a$, and significant duration $D_{5-95}$  }
\label{sec6_fig1} 
\end{figure}

This subsection validates the hierarchical GMMs by comparing four key IMs: PGA, PGV, Arias intensity $I_a$, and significant duration $D_{5-95}$.  The results are shown in Figure \ref{sec6_fig1}, which compares empirical CDFs of the four IMs computed from the real (blue lines) and synthetic datasets (grey lines), with the red dashed lines denoting the mean of the synthetic datasets.  The results indicate that both joint PDF models are capable of generating synthetic datasets that closely resemble the real data, with most synthetic realizations falling within the 2-$\sigma$ confident bounds (red dotted lines). The most significant deviation pertains to PGV, and this is inherited from the baseline GMM, as shown in Figure \ref{sec4_fig0}. Observe that the additional randomness in the gray lines is given by the uncertainties in the GMM parameters. 


\subsection{Validation of linear-response spectra}
\label{sec6_1}
\begin{figure}[!htb]
   \centering
   \includegraphics[width=0.67\textwidth]{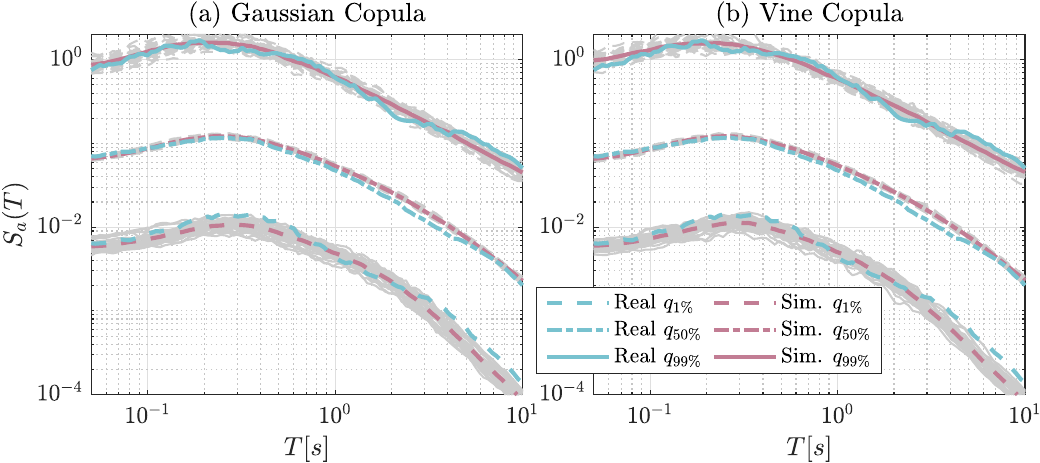}
\caption{Validation of the hierarchical GMM using two different copula models w.r.t. three quantiles of 5\%-damped linear-response spectra.}
\label{sec6_fig2} 
\end{figure}

Next, we compare the 5\%-damped linear-response spectra $Sa(T)$ between the real and synthetic datasets. In Figure \ref{sec6_fig2}, a comparison is made between three quantiles ($q_{1\%}$, $q_{50\%}$, and $q_{99\%}$) of $Sa(T)$ computed from the real (blue lines) and synthetic datasets (grey lines). The red lines represent the average results of their corresponding 30 grey lines. As before, the regions encompassed by the grey lines are viewed as confidence bounds of the red lines. 

The figure demonstrates that both PDF models offer a similar performance. In terms of matching the red lines to the blue lines, both models show accurate agreement for the engineering meaningful  $q_{99\%}$ curves. Moreover there is a good matching for the $q_{50\%}$ curves in the $[0.05, 2]$s range. However, biases are noticeable in other regions, with simulations tending to overestimate the $q_{50\%}$ curves and underestimate the $q_{1\%}$ curves. Most biases remain within the confidence bounds, except for the $q_{50\%}$ curves in the $[1.5, 5]$s range, where deviations are more pronounced.
Moreover, Figure \ref{sec6_fig2} shows that the two copula models do not yield significant differences in various spectral quantiles. 

\begin{figure}[!htb]
   \centering
   \includegraphics[width=0.67\textwidth]{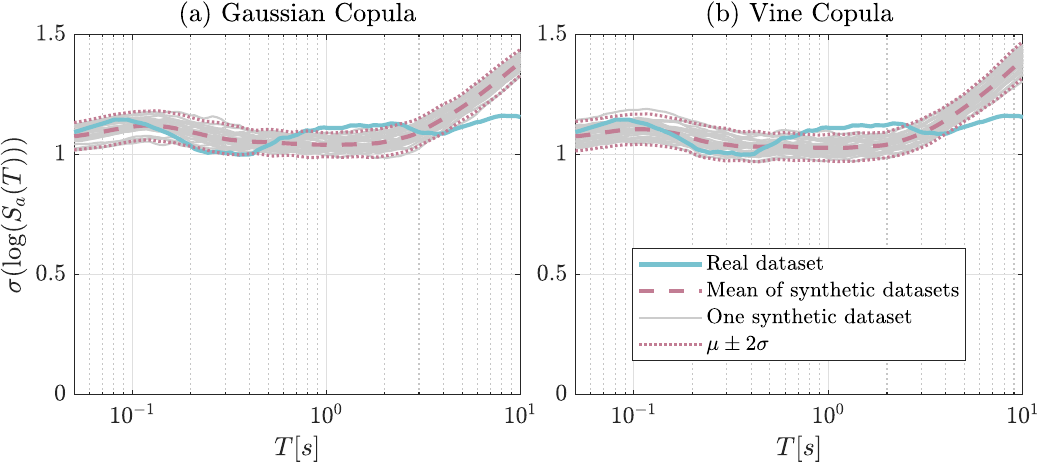}
\caption{Validation of the hierarchical GMM using two different copula models w.r.t. logarithmic standard deviation of 5\%-damped linear-response spectra.}
\label{sec6_fig3} 
\end{figure}

Next, Figure \ref{sec6_fig3} compares the standard deviation of $\log(S_a(T))$ computed from the real and synthetic GM datasets. The figure illustrates that, regardless of the PDF model, the mean simulation results (red dashed lines) generally align with the target recorded values (blue lines), which fall within the $\mu \pm 2\sigma$ confidence bounds  (computed from the grey lines). Compared to  Figure \ref{sec4_fig2} with $\zeta = {5\%}$, the confidence bounds in Figure \ref{sec6_fig3} are wider. This suggests that the two sources of randomness inside the hierarchical GMM, namely $\boldsymbol{\Theta}$ and $\boldsymbol{Z}$,  both contribute to the variability of  the predicted spectral variability, with the former making a slightly larger impact.

\begin{figure}[!htb]
   \centering
   \includegraphics[width=0.98\textwidth]{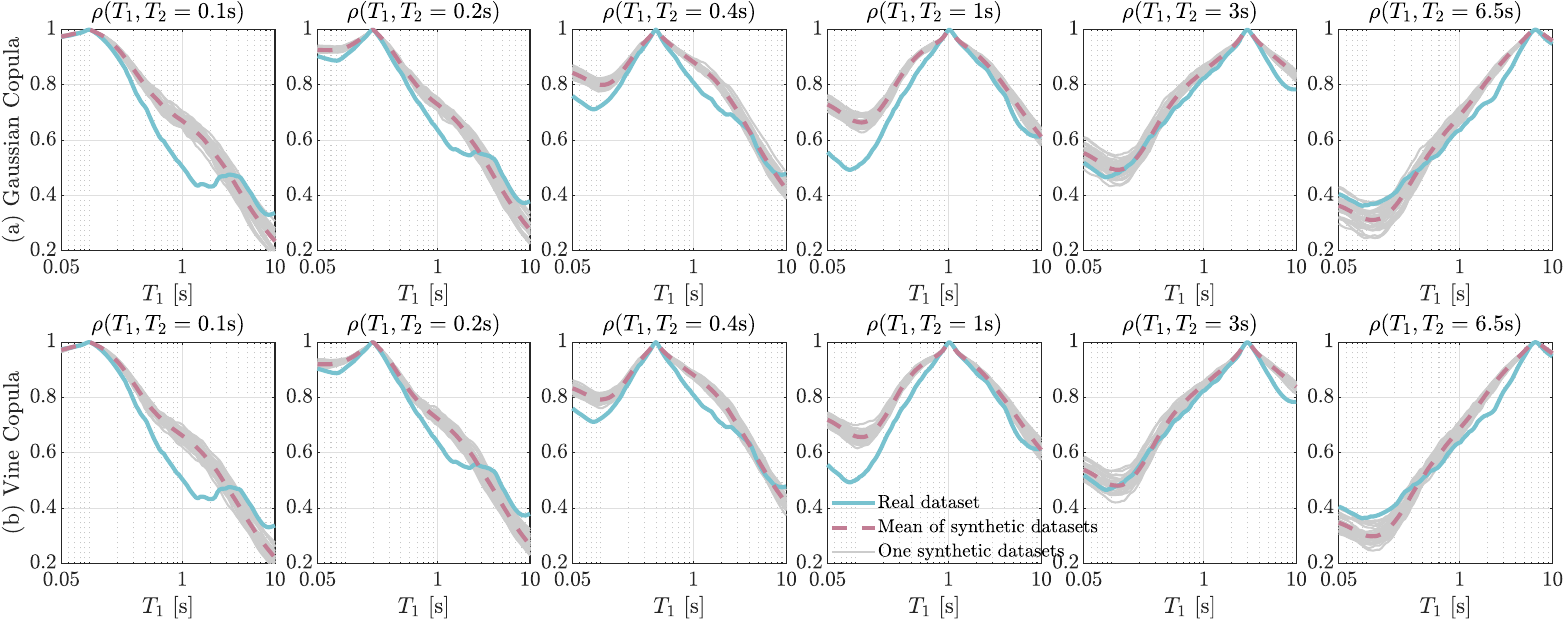}
\caption{Validation of the hierarchical GMM using two different copula models w.r.t. the spectral correlation of 5\%-damped linear-response spectra.}
\label{sec6_fig4} 
\end{figure}

Finally, Figure \ref{sec6_fig4} compares Pearson's correlation coefficients, $\rho(T_1,T_2)$, for the response spectrum $Sa(T)$ between $T_1 \in [0.05, 10] $s and six fixed periods $T_2$. Both the PDF models effectively capture the correlation trends of the real dataset across most period ranges. The predicted correlations in Figure \ref{sec6_fig4} are similar to the results obtained using the real GMM parameters (shown in Figure \ref{sec4_fig3} (b)), with two key differences: (1) the confidence bounds in Figure \ref{sec6_fig4} are wider, and (2) the synthetic dataset mean more closely matches the real dataset, particularly for the curve with  $T_2 = 3 $s.

\subsection{Validation of nonlinear-response spectra}
\label{sec6_2}
\begin{figure}[!htb]
   \centering
   \includegraphics[width=0.67\textwidth]{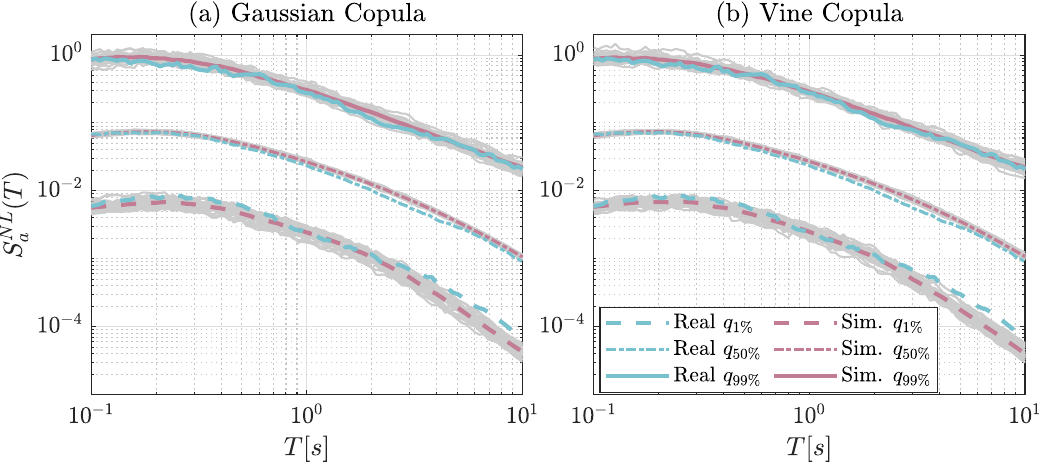}
\caption{Validation of the hierarchical GMM using two different copula models w.r.t. three quantiles of constant-ductility ($\mu=2$) nonlinear-response spectra.}
\label{sec6_fig5} 
\end{figure}

\begin{figure}[!htb]
   \centering
   \includegraphics[width=0.67\textwidth]{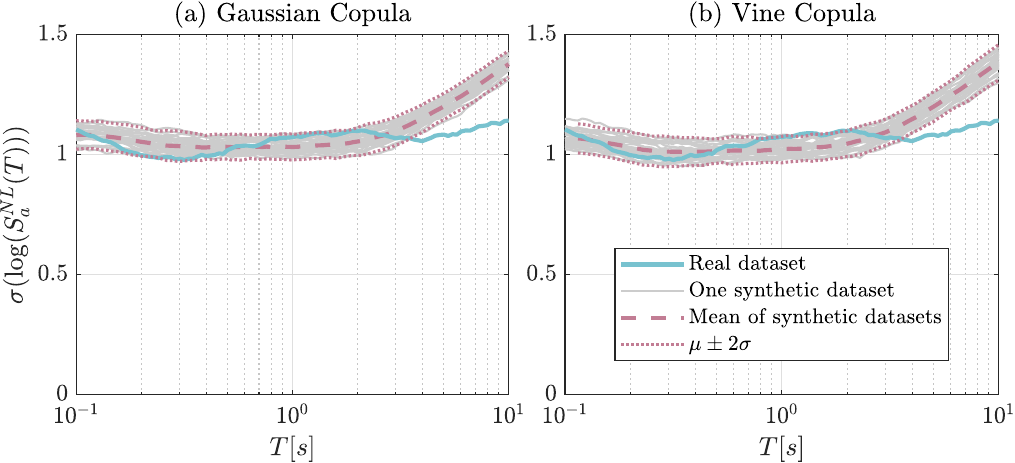}
\caption{Validation of the hierarchical GMMs using two different copula models w.r.t. logarithmic standard deviation of constant-ductility ($\mu=2$) nonlinear-response spectra.}
\label{sec6_fig6} 
\end{figure}

\begin{figure}[!htb]
   \centering
   \includegraphics[width=0.98\textwidth]{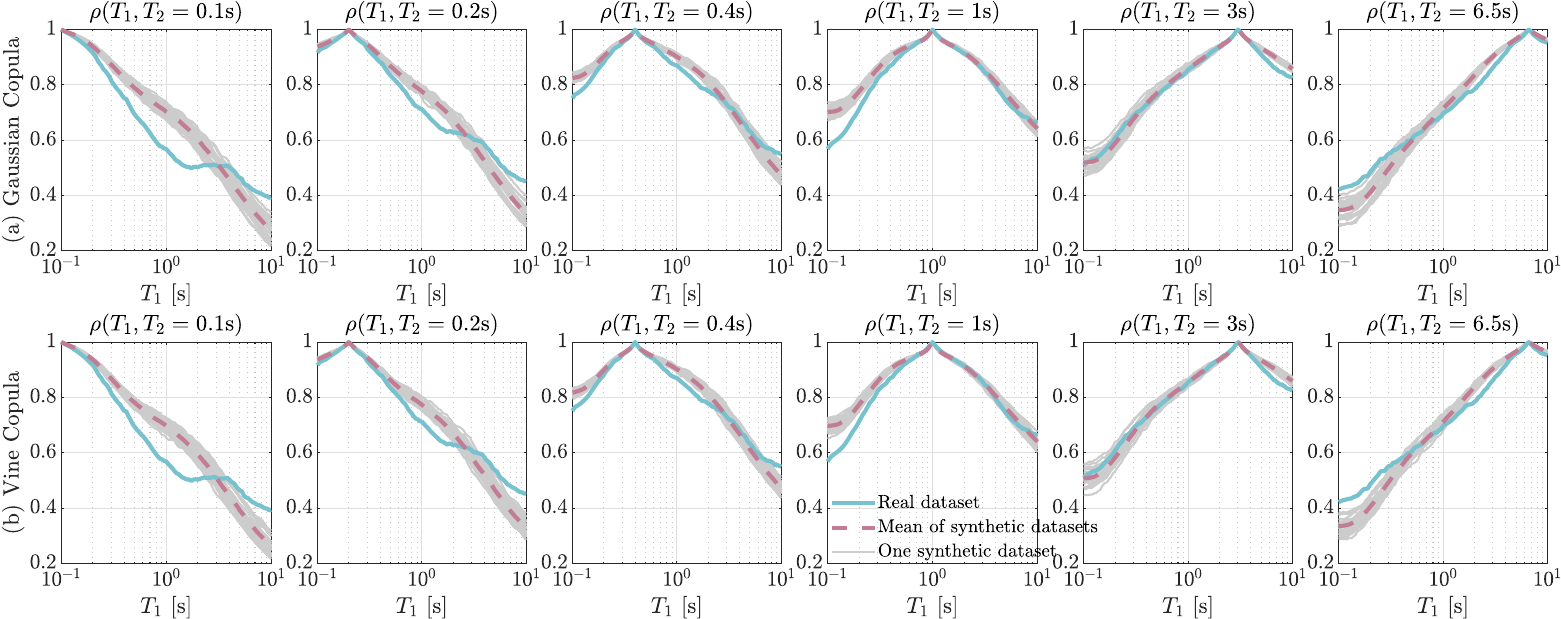}
\caption{Validation of the hierarchical GMMs using two different copula models w.r.t. spectral correlation of constant-ductility ($\mu=2$) nonlinear-response spectra.}
\label{sec6_fig7} 
\end{figure}

This section validates the constant-ductility nonlinear-response spectra $S^{NL}_a(T)$, with the ductility factor $\mu$ set to 2. The same comparison method used in the previous subsection is adopted. The comparison of statistical quantities includes three quantiles ($q_{1\%}$, $q_{50\%}$, and $q_{99\%}$) of $S^{NL}_a(T)$ (in Figure \ref{sec6_fig5}), the logarithmic standard deviation of $\log(S_a^{NL}(T))$ (in Figure \ref{sec6_fig6}), and the nonlinear spectral correlation  (in Figure \ref{sec6_fig7}). These figures indicate that the results of  $S^{NL}_a(T)$ are similar to the results of the linear-response spectra. Specifically, regardless of the choice of copula model, the synthetic GM datasets well match the real GM dataset in terms of  nonlinear-response spectra statistics.

\subsection{Validation using train-test split}
\label{sec6_4}
\begin{figure}[htb]
\centering
\begin{subfigure}[b]{0.45\textwidth}
   \includegraphics[width=1\linewidth]{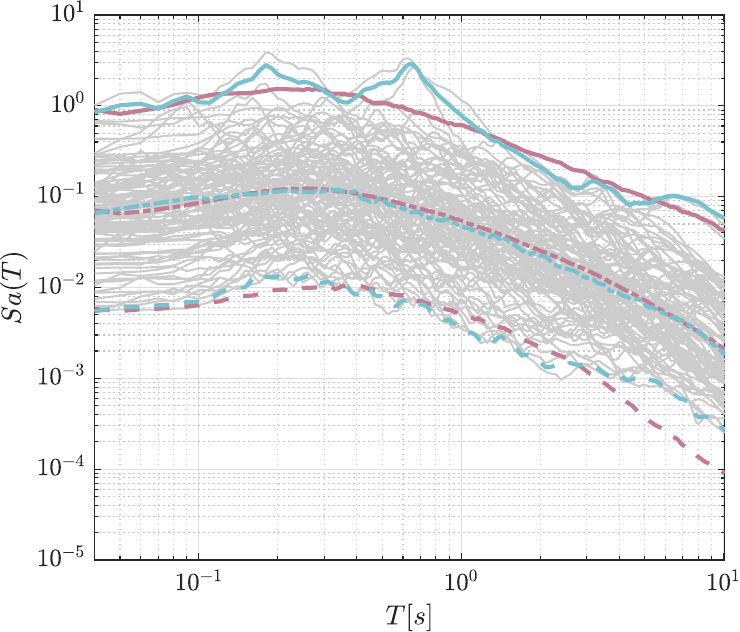}
   \caption{linear-response spectrum $S_a(T|\zeta=5\%)$}
   \label{sec6_fig8a} 
\end{subfigure}
 \hfill
\begin{subfigure}[b]{0.45\textwidth}
   \includegraphics[width=1\linewidth]{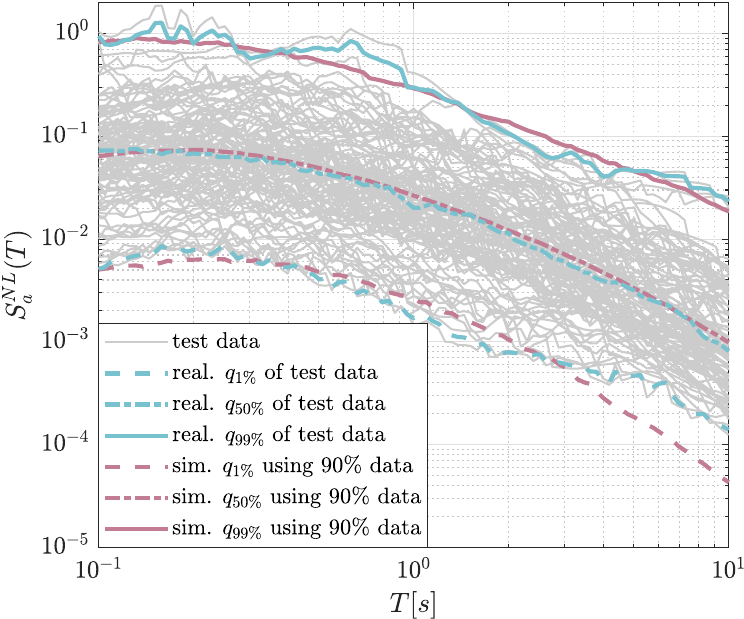}
   \caption{nonlinear-response spectrum $S_a^{NL}(T|\mu=2)$}
   \label{sec6_fig8b} 
\end{subfigure}
\caption{Validation of the hierarchical GMMs using train-test split w.r.t. (a) three quantiles of 5\%-damped linear-response spectra and (b) three quantiles of constant-ductility ($\mu=2$) nonlinear-response spectra.}
\label{sec6_fig8} 
\end{figure}

\begin{figure}[htb]
\centering
\begin{subfigure}[b]{0.98\textwidth}
   \includegraphics[width=1\linewidth]{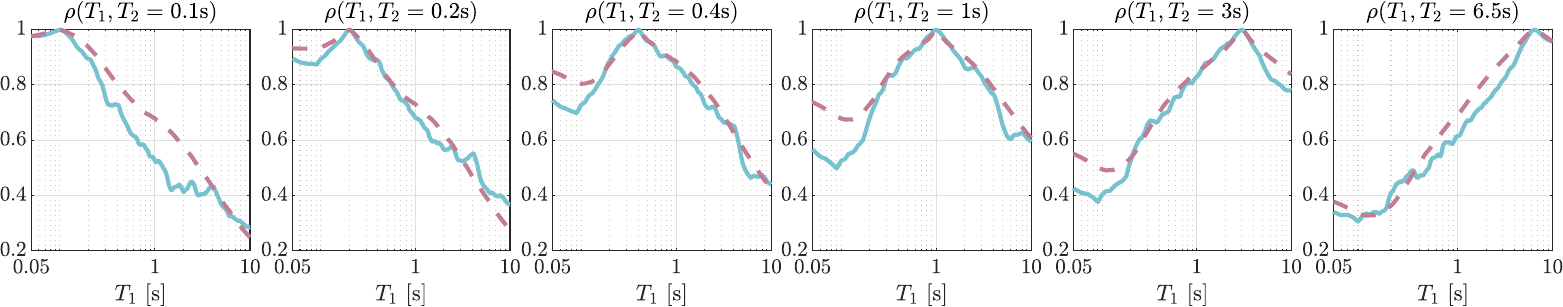}
   \caption{linear-response spectrum $S_a(T|\zeta=5\%)$}
   \label{sec6_fig9a} 
\end{subfigure}
 \hfill
\begin{subfigure}[b]{0.98\textwidth}
   \includegraphics[width=1\linewidth]{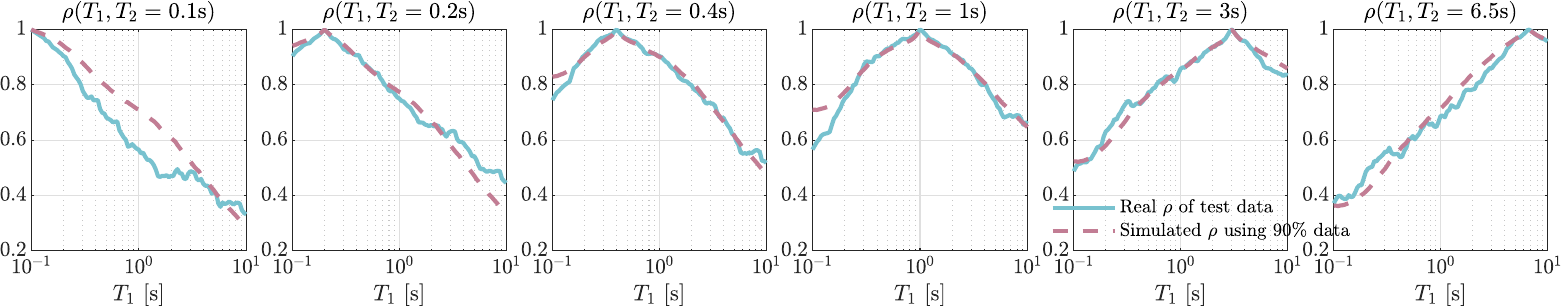}
   \caption{nonlinear-response spectrum $S_a^{NL}(T|\mu=2)$}
   \label{sec6_fig9b} 
\end{subfigure}
\caption{Validation of the hierarchical GMMs using train-test split w.r.t. (a) spectral correlation $\rho(T_1,T_2)$ of 5\%-damped linear-response spectra and (b) spectral correlation $\rho(T_1,T_2)$ of constant-ductility ($\mu=2$) nonlinear-response spectra.}
\label{sec6_fig9} 
\end{figure}

To evaluate model generalization, we randomly split the 1,001-GM dataset into a 90\% training set and a 10\% test set. The training set is used to fit a hierarchical GMM, following the procedure outlined in this study. This hierarchical GMM combines the baseline stochastic GMM with a Gaussian-copula-based joint PDF. After training, we compare the statistics of response spectra computed from  5,000 samples drawn from the hierarchical GMM with those from test set. These statistics include the spectral quantities in Figure \ref{sec6_fig8} and the spectral correlation in Figure  \ref{sec6_fig9}.  
In Figure  \ref{sec6_fig8},  the simulated quantiles closely track the real quantiles across a wide range of periods for both the linear and nonlinear response spectra. The largest biases occur at the 1\% quantile for long periods. However, this bias has less engineering significance, because higher-amplitude responses are more critical for engineering applications. In Figure  \ref{sec6_fig9}, the simulated spectral correlations align well with the real correlations for most period pairs ($T_1,T_2$). The most noticeable bias occurs at $T_1=0.1$s, where the simulated motions tend to overestimate the real correlation at middle periods. This issue is traced back to the limitations in the baseline model, as illustrated in Figure \ref{sec4_fig3} (b). 

In summary, the hierarchical model, trained on a reduced dataset, accurately predicts the statistics of the unseen test set, indicating that the training and test sets follow similar probabilistic distributions. This supports the ergodicity assumption, suggesting that strong-motion datasets from different regions share a similar distribution, allowing us to use global data to construct stochastic GMMs for predicting future GM time series. Further studies are required to rigorously test this ergodicity assumption.

\section{Discussion and Conclusions}
\subsection{Discussion}
The determination of the optimal GMM depends on the selected validation metrics, ranking criteria, and the selected dataset. As a result, the ``optimal" model can vary when different metrics or criteria are applied. Additionally, some datasets may include ground motions with unique or atypical characteristics, such as recorded time series with significant spikes or pulse-like features. In such cases (considered as outliers), it may be necessary to modify or enhance the model configurations to account for these characteristics or to refine the dataset.

This study focused on validating the GMs with respect to classical IMs. However, there is a second layer of validation—beyond the scope of this work—that involves validating structural demands/responses. In this context, a complete GMM validation requires comparison metrics that cover a comprehensive range of structural demands/responses of interest. However, this becomes computationally prohibitive when complex models are involved, presenting a persistent challenge for both researchers and engineers. As noted in \cite{rezaeian_findings_2023}, the validation process is application-specific, meaning the metrics must be tailored to the specific engineering or scientific applications. In this study, the (elastic and inelastic) response spectrum represents the first layer of validation with respect to structural demands for SDOF systems. Our future work will focus on comparing simulated GMs with recorded data using validation metrics tailored to broader structural applications and structural demands/responses.

Selecting the appropriate copula model requires more extreme data for calibration. Although this paper finds no significant difference between using the Gaussian copula and the R-vine copula, the R-vine copula may offer greater flexibility in capturing tail dependence. R-vine copula is a powerful tool for robustly simulating extreme seismic events (i.e., simulating within the tails of the joint PDF of the GMM parameters). Further research with larger datasets and extreme values is necessary to fully evaluate the potential benefits of using the R-vine copula.

The engineering applications of hierarchical stochastic GMMs are twofold. 
The first is the \textit{data-driven stochastic hierarchical GMM}, which fits a specific dataset by leveraging records from various locations and time periods. This approach relies on the ergodicity assumption, i.e., long-term temporal variability at a single site can be represented by observations from different sites (spatial variability) that share similar seismological properties with the site of interest. Moreover, a ``small" dataset of selected GMs can be augmented through simulations, providing an unlimited supply of motions that share the characteristics of the selected ones. This approach can thus be integrated with ground motion selection processes, effectively enabling the application of classical forward uncertainty quantification (UQ) methods in this field. Notably, this also provides the opportunity to validate the results obtained from UQ methods using real GMs.

The second application is the \textit{site-based stochastic GMM}, similar to \cite{rezaeian_simulation_2012,vlachos_predictive_2018,dabaghi_simulation_2018}. In this method, predictive equations are developed to relate GMM parameters to specific seismological variables (e.g., $M$, $R$, $V_{S30}$). This approach enables the use of hierarchical models to generate a family of potential GMs for specified earthquake scenarios, specifically in cases where sufficient records are lacking. This framework is inherently UQ-based, allowing for the exploration of various research questions. These include sensitivity analysis with respect to the different types of variables (i.e., site-specific variables vs. ground-motion-specific variables) and the development of efficient stochastic surrogate models for complex structures \cite{zhu_seismic_2023}. Additionally, the predictive equations can be reformulated as joint PDFs between site and GM variables, enabling all desired conditional simulations (e.g., extreme events). Our future work will focus in this topic.

\subsection{Summary and Conclusions}

This paper presents an essential review/collection of hierarchical stochastic ground motion models (GMMs) for simulating nonpulse-like time series, restructured within a modular framework. The GMMs incorporate two layers of uncertainties: (1) model's built-in white noise, capturing  the intrinsic variability of individual ground motions (GMs), and (2) random GMM parameters, such as Arias intensity, predominate frequency, and motion duration, representing record-to-record variability.
 
The hierarchical stochastic GMMs are developed in two steps:  (1) selecting a modulated filtered white noise model  (MFWNM) and (2) building a joint  probability density function (PDF) for the GMM parameters.
In the first step, the MFWNM is formulated using spectral representation, offering flexibility in  modeling time-varying spectral content. We specifically  explore  different frequency filter types (single- and multi-mode) and various trend functions (constant, linear, and non-parametric) to describe the filters' time-varying properties.  
In the second step, using Sklar's theorem, the joint PDF is decomposed into a product of marginal distributions and a dependence structure, represented by copula models. We explore two copula models, the Gaussian copula and the R-vine copula, to evaluate their impact on the characteristics of the generated GMs. 
Two key findings of the study are: 
\begin{itemize}
    \item  \textbf{Hierarchical stochastic GMMs are validated on a large dataset}.  Hierarchical models generate synthetic GM datasets that are statistically compatible with a large dataset comprising 1,001 strong-motion records.   The compatible statistics include the cumulative distribution functions of four intensity measures (peak ground acceleration (PGA), peak ground velocity (PGV), Arias intensity, and significant duration), as well as quantiles, standard deviations, correlations of both elastic- and inelastic-response spectra.
    \item  \textbf{A parsimonious hierarchical stochastic GMM is identified}. Based on the selected validation metrics and ranking strategy, an 11-parameter MFWNM, integrating  a single-mode second-order filter and simple trend functions, offers sufficient accuracy.  Increasing the complexity of the MFWNM to 33 parameters to account for more complex time-varying spectral content yields minimal improvement. Furthermore, the more complex R-vine copula model performs similarly to the widely used Gaussian copula model. Therefore, for far-filed non-pulse-like ground motion simulations, we recommend the simple 11-parameter model (i.e., Model 1) with a Gaussian copula. 
\end{itemize}

\section*{Acknowledgments}
The first authors are funded by the Italian Ministry of Education, University and Research (MIUR) in the frame of the ``Departments of Excellence 2023-2027'' (grant L 232/2016). The last author is by funded the European Union under Next GenerationEU. PRIN 2022 Prot. n 2022MJ82MC\_001.

\appendix
\counterwithin{figure}{section} 
\renewcommand{\thefigure}{A\arabic{figure}} 
\setcounter{figure}{0} 

\section{Ranking results in the optimal GMM study}
\label{appendix_2}
\begin{figure}[!htb]
   \centering
   \includegraphics[width=0.98\textwidth]{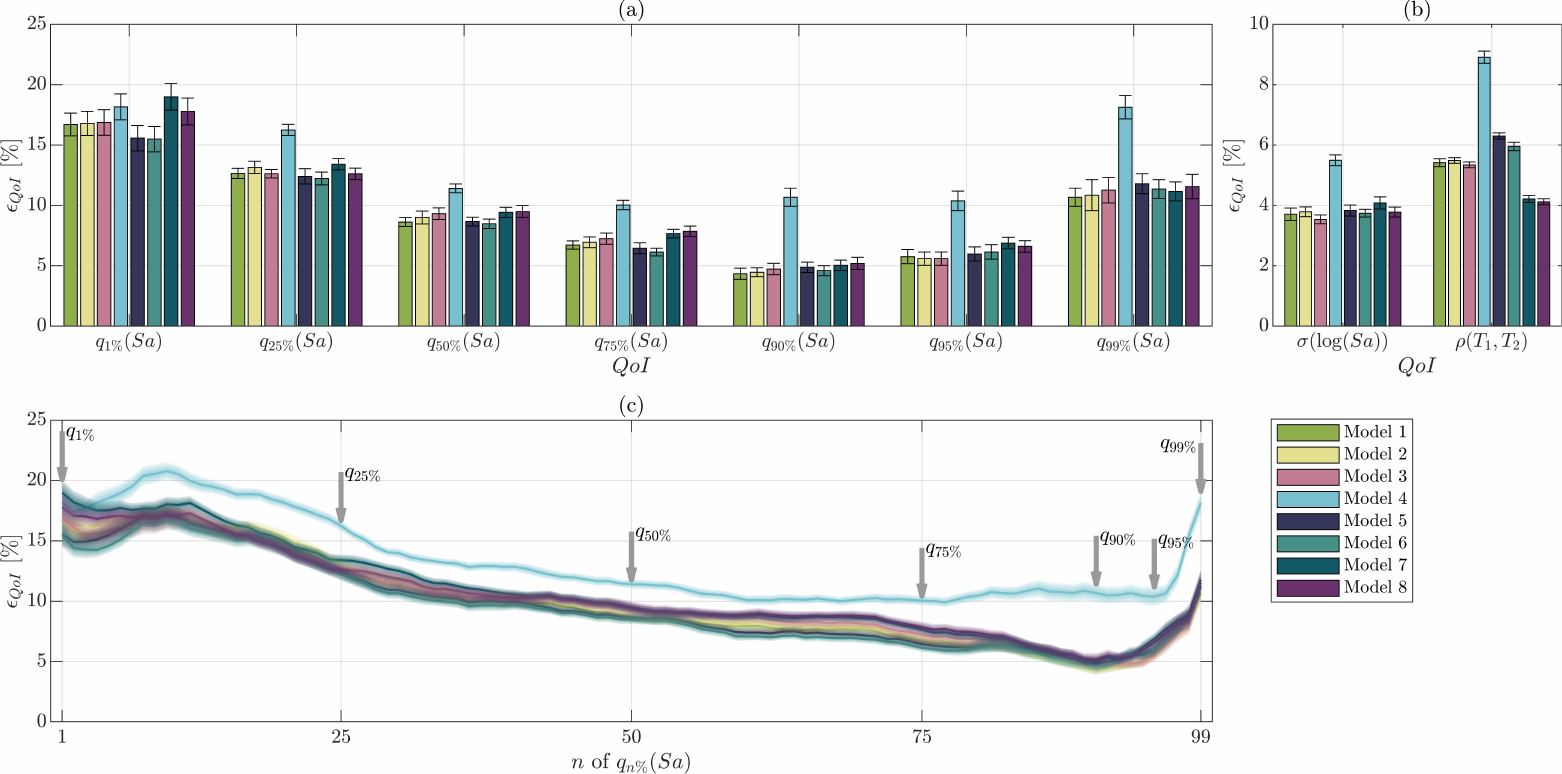}
\caption{Compare error metrics $\epsilon_{QoI}$ of 2\%-damped $Sa(T)$ for the eight considered parametric GMMs. (a) and (b): Error bars of different metrics $\epsilon_{QoI}$, with the whiskers representing $\pm 1\sigma$ confidence bounds; (c): Biases of $q_{n\%}$ vary with the quantile level $n$,  with $\pm 1\sigma$ confidence bounds.}
\label{sec4_fig6} 
\end{figure}
\begin{figure}[!htb]
   \centering
   \includegraphics[width=0.98\textwidth]{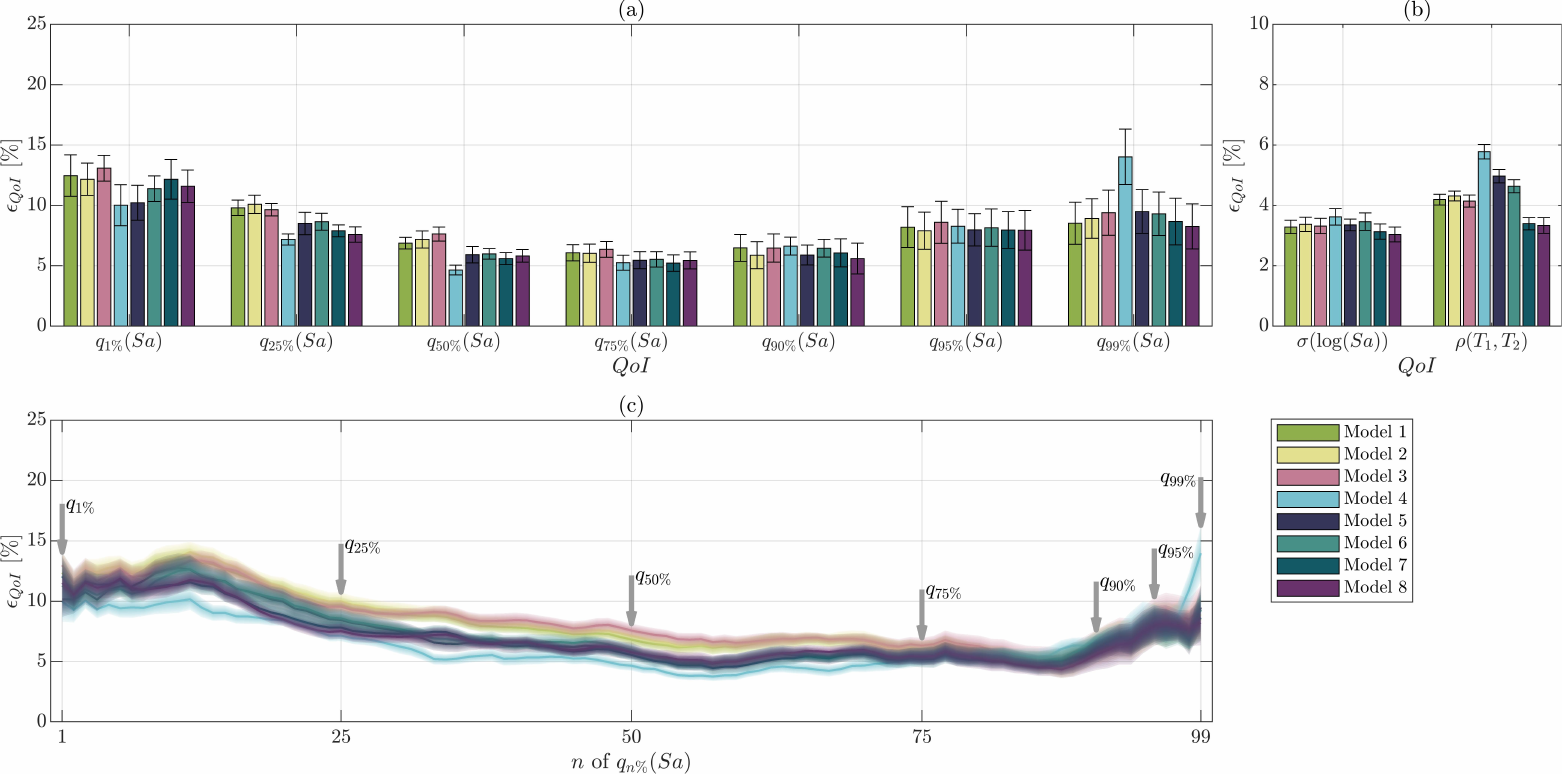}
\caption{Compare error metrics $\epsilon_{QoI}$  of 20\%-damped $Sa(T)$ for the eight considered parametric GMMs. (a) and (b): Error bars of different metrics $\epsilon_{QoI}$, with the whiskers representing $\pm 1\sigma$ confidence bounds; (c): Biases of $q_{n\%}$ vary with the quantile level $n$,  with $\pm 1\sigma$ confidence bounds.}
\label{sec4_fig8} 
\end{figure}

\begin{figure}[!htb]
   \centering
   \includegraphics[width=0.98\textwidth]{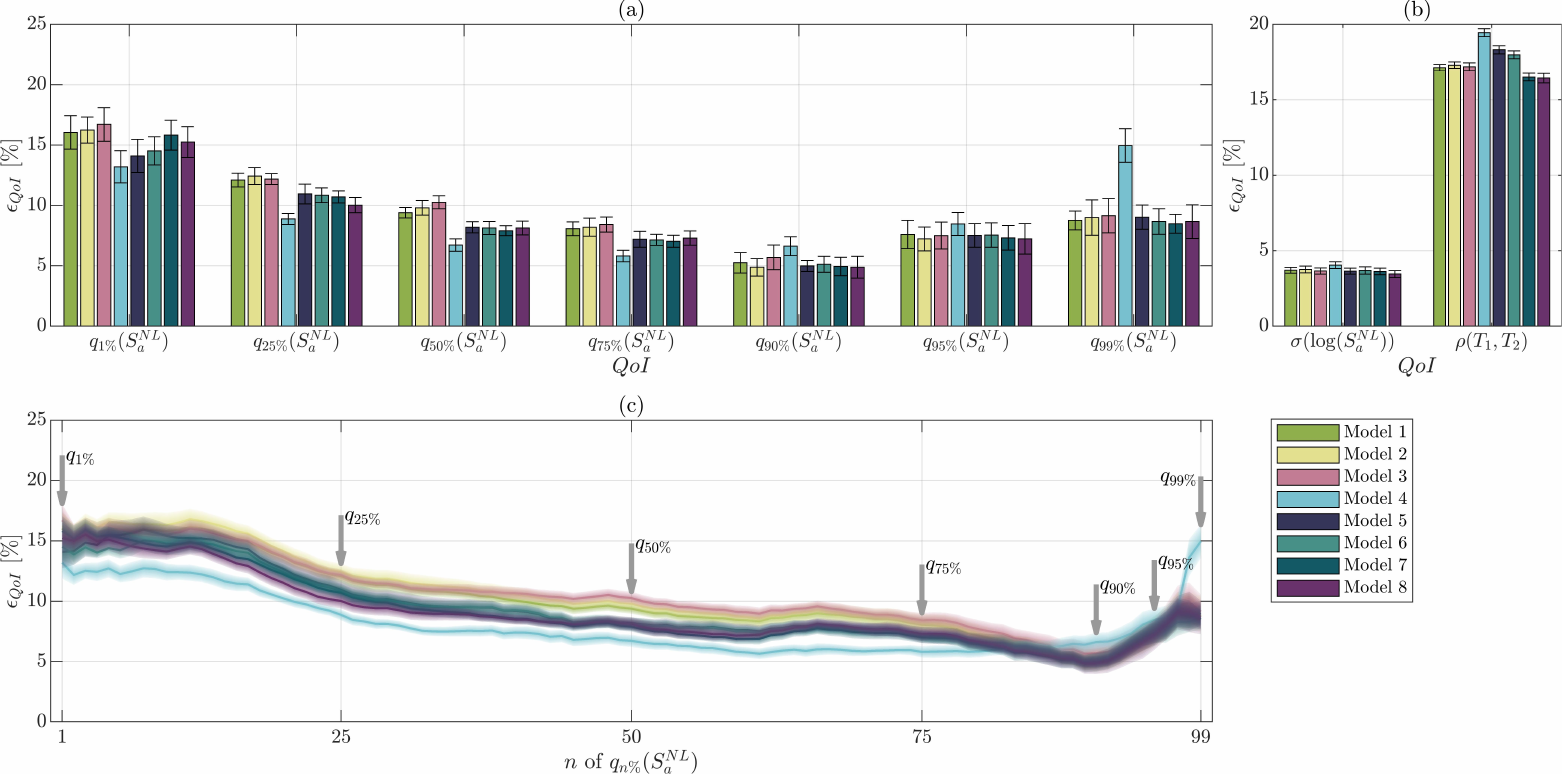}
\caption{Compare error metrics $\epsilon_{QoI}$  of constant-ductility ($\mu=1.5$) spectra for the eight considered parametric GMMs. (a) and (b): Error bars of different metrics $\epsilon_{QoI}$, with the whiskers representing $\pm 1\sigma$ confidence bounds; (c): Biases of $q_{n\%}$ vary with the quantile level $n$,  with $\pm 1\sigma$ confidence bounds.}
\label{sec4_fig9} 
\end{figure}
\begin{figure}[!htb]
   \centering
   \includegraphics[width=0.98\textwidth]{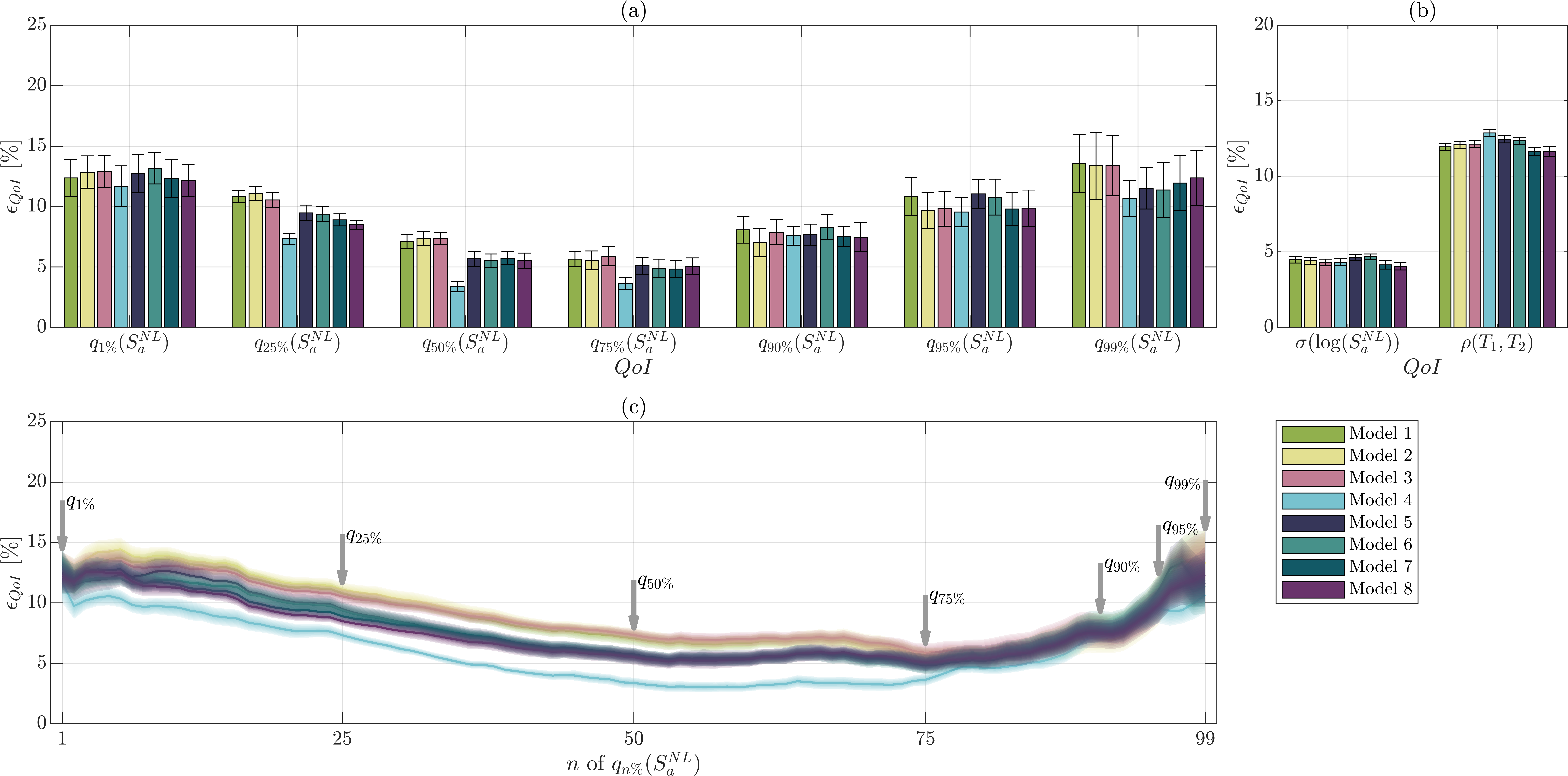}
\caption{Compare error metrics $\epsilon_{QoI}$ of constant-ductility ($\mu=4$) spectra for the eight considered parametric GMMs. (a) and (b): Error bars of different metrics $\epsilon_{QoI}$, with the whiskers representing $\pm 1\sigma$ confidence bounds; (c): Biases of $q_{n\%}$ vary with the quantile level $n$,  with $\pm 1\sigma$ confidence bounds.}
\label{sec4_fig11} 
\end{figure}

\clearpage
\bibliographystyle{elsarticle-num} 
\bibliography{MyRef} 

\end{document}